%% file: CSV19.tex
\documentclass[runningheads]{llncs}
\pdfoutput=1
\usepackage{amssymb}
\usepackage{amsmath}
\usepackage{epsfig}
\usepackage{graphicx} 
\usepackage{graphics, color}
\usepackage{hyperref}
\usepackage{algorithm}
\usepackage{algpseudocode}
\usepackage{url}
\newfloat{algorithm}{t}{lop}
\usepackage{etoolbox}
\usepackage{verbatim}
\usepackage{listings}
\usepackage{tabularx}
\usepackage{multirow}
\usepackage{subcaption}
\usepackage{bbm}
\usepackage{times}
\usepackage{xcolor}
\usepackage{soul}
\usepackage[utf8]{inputenc}

\newcommand{\mat}[2][]{%
	\ifstrempty{#1}{%
		\boldsymbol{#2}
	}{%
	\boldsymbol{#1}^{[#2]}
}%
}
\newcommand{\npm}[2][]{%
	\#PM(#2)
}%

\newcommand{\perm}[2][]{%
	perm(#2)
}%

\newcommand{\radd}{\ensuremath{\mathsf{RysersADD}}}
\newcommand{\raddp}{\ensuremath{\mathsf{RysersADD{\text -}P}}}

\newcommand{\dfr}{\ensuremath{\mathsf{D4}}}
\newcommand{\ds}{\ensuremath{\mathsf{DSharp}}}

\newcommand{\clusterrank}{\ensuremath{\mathsf{clusterRank}}}
\newcommand{\addmc}{\ensuremath{\mathsf{ADDMC}}}
\newcommand{\cvo}{\ensuremath{\eta}}

\algnotext{EndFor}
\algnotext{EndIf}
\algnotext{EndWhile}
\algnotext{Until}

\title{On Symbolic Approaches for Computing the Matrix Permanent\thanks{Author names are ordered alphabetically by last name and does not indicate contribution}\thanks{Work supported in part by NSF grant IIS-1527668, the Data Analysis and Visualization Cyberinfrastructure funded by NSF under grant OCI-0959097 and Rice University, and MHRD IMPRINT-1 Project No. 6537 sponsored by Govt of India. Authors are grateful to Vu Phan and Jeffrey Dudek for help with ADDMC.}\thanks{This is a post-peer-review, pre-copyedit version of an article to be published in the proceedings of CP'19.  The link to the final authenticated version will be updated once available.}}
\author{Supratik Chakraborty\inst{1} \and
	Aditya A. Shrotri\inst{2} \and
	Moshe Y. Vardi\inst{2}}
\institute{Indian Institute of Technology Bombay, India 
	\email{supratik@cse.iitb.ac.in} \and
	Rice University, Houston, USA
	\email{\{Aditya.Aniruddh.Shrotri,vardi\}@rice.edu}}

\begin{document}
	
	\maketitle
	\input{Abstract}
	\input{Introduction}
	\input{Notation}
	\input{Related}
	\input{Algorithm}
	\input{Experiments}

	\input{Conclusion}
	\bibliographystyle{abbrv}
	\bibliography{Refs}
	
\end{document}

%% file: Abstract.tex
\begin{abstract}
Counting the number of perfect matchings in bipartite graphs, or
equivalently computing the permanent of $0$-$1$ matrices, is an
important combinatorial problem that has been extensively studied by
theoreticians and practitioners alike. The permanent is \#P-Complete; hence
it is unlikely that a polynomial-time algorithm exists for the
problem. 
Researchers have therefore focused on finding
tractable subclasses of matrices for permanent computation.
One such subclass that has received
much attention 
is that of sparse matrices i.e. matrices with few entries set to $1$,
the rest being $0$. For this subclass, improved theoretical upper
bounds and practically efficient algorithms have been developed. In
this paper, we ask whether it is possible to go beyond sparse matrices
in our quest for developing scalable techniques for the permanent,
%
and answer this question affirmatively. Our key insight is to
represent permanent computation symbolically using Algebraic Decision
Diagrams (ADDs). ADD-based techniques naturally use dynamic
programming, and hence avoid redundant computation through
memoization. This permits exploiting the hidden structure in a large
class of matrices that have so far remained beyond the reach of
permanent computation techniques. The availability of sophisticated
libraries implementing ADDs also makes the task of engineering
practical solutions relatively straightforward. While a complete
characterization of matrices admitting a compact ADD representation
remains open, we provide strong experimental evidence of the
effectiveness of our approach for computing the permanent, not just
for sparse matrices, but also for dense matrices and for matrices with
``similar'' rows.
\end{abstract}

%% file: Introduction.tex
\section{Introduction}

Constrained counting lies at the heart of several important problems in diverse areas such as performing Bayesian inference~\cite{Roth1996}, measuring resilience of electrical networks~\cite{duenas2017counting}, counting Kekule structures in chemistry~\cite{gordon1952theory}, computing the partition function of monomer-dimer systems~\cite{huo2008computing}, and the like.  Many of these problems reduce to counting problems on graphs. For instance, learning probabilistic models from data reduces to counting the number of topological sorts of directed acyclic graphs~\cite{wallace1996causal}, while computing the partition function of a monomer-dimer system reduces to computing the number of perfect matchings of an appropriately defined bipartite graph~\cite{huo2008computing}.  In this paper, we focus on the last class of problems -- that of counting perfect matchings in bipartite graphs.  It is well known that this problem is equivalent to computing the \emph{permanent} of the $0$-$1$ bi-adjacency matrix of the bipartite graph. We refer to these two problems interchangeably in the remainder of the paper.

Given an $n\times n$ matrix $ \mat{A} $ with real-valued entries, the permanent of $\mat{A}$ is given by
$\perm{\mat{A}} = \sum_{\sigma\in S_n} \prod_{i=1}^{n} a_{i,\sigma(i)}$,
where $S_n$ denotes the symmetric group of all permutations of $1,
\ldots n$.
This expression is almost identical to that for the determinant of
$\mat{A}$; the only difference is that the determinant includes the sign of the permutation in the inner product. Despite the striking resemblance of the two expressions, the complexities of computing the permanent and determinant are 
vastly different. While the determinant can be computed in time
$\mathcal{O}(n^{2.4})$, Valiant~\cite{Valiant79} showed that computing
the permanent of a $0$-$1$ matrix is \#P-Complete, making a polynomial-time algorithm unlikely~\cite{Toda89}. Further evidence of the hardness of computing
the permanent was provided by Cai, Pavan and
Sivakumar~\cite{cai1999hardness}, who showed that the permanent is
also hard to compute on average. Dell et
al.~\cite{dell2014exponential} showed that there can be no algorithm
with sub-exponential time complexity, assuming a weak version of the
Exponential Time Hypothesis~\cite{AroBar09} holds.

The determinant has a nice geometric interpretation: it is the
oriented volume of the parallelepiped spanned by the rows of the
matrix.  The permanent, however, has no simple geometric
interpretation. Yet, it finds applications in a wide range of areas.
In chemistry, the permanent and the
permanental polynomial of the adjacency matrices of fullerenes~\cite{kroto1985c60} have attracted much attention over the
years~\cite{cash1995fast,liang2004partially,chou2015computing}. In constraint programming, solutions to All-Different constraints can be expressed as perfect matchings in a bipartite graph~\cite{regin1994filtering}. An estimate of the number of such solutions can be used as a branching heuristic to guide search~\cite{zanarini2009solution,pesant2012counting}. In physics, permanents can be used to measure quantum entanglement~\cite{wei2010matrix} and to compute the partition functions of monomer-dimer systems~\cite{huo2008computing}. 


Since computing the permanent is hard in general, researchers have
attempted to find efficient solutions for either approximate versions
of the problem, or for restricted classes of inputs.  In this paper,
we restrict our attention to exact algorithms for computing the
permanent. The asymptotically fastest known exact algorithm for general $n\times n$
matrices is Nijenhuis and Wilf's version of Ryser's
algorithm~\cite{ryser1963combinatorial,nijenhuis2014combinatorial},
which runs in time ${\Theta}(n\cdot2^n)$ for all matrices of size $ n $. For matrices with bounded treewidth or clique-width~\cite{robertson1984graph,courcelle1993handle}, Courcelle,
Makowsky and Rotics~\cite{courcelle2001fixed} showed that the
permanent can be computed in time linear in the size of the matrix,
i.e., computing the permanent is Fixed Parameter Tractable (FPT).  A
large body of work is devoted to developing fast algorithms for sparse
matrices, i.e. matrices with only a few entries set to non-zero
values~\cite{servedio2005computing,liang2004partially,izumi2012new,yue2013improved}
in each row.  Note that the problem remains \#P-Complete
even when the input is restricted to matrices with exactly three $1$'s per row and column~\cite{broder1986hard}.

An interesting question to ask is whether we can go beyond sparse
matrices in our quest for practically efficient algorithms for the
permanent.  For example, can we hope for practically efficient
algorithms for computing the permanent of \emph{dense} matrices, i.e.,
matrices with almost all entries non-zero?  Can we expect efficiency
when the rows of the matrix are ``similar'', i.e.  each row has only a
few elements different from any other row (sparse and dense matrices
being special cases)?  Existing results do not seem to throw much
light on these questions.  For instance, while certain non-sparse
matrices indeed have bounded clique-width, the aforementioned result of
Courcelle et al~\cite{courcelle1990graph,courcelle2001fixed} does not
yield practically efficient algorithms as the constants involved are enormous~\cite{grohe1999descriptive}. The hardness of non-sparse instances is underscored by the fact that SAT-based model counters do not scale well on these, despite the fact that years of research and careful engineering have enabled these tools to scale extremely well on a diverse array of problems. We experimented with a variety of CNF-encodings of the permanent on state-of-the-art counters like {\dfr}~\cite{lagniez2017improved}. Strikingly, no combination of tool and encoding was able to scale to matrices even half the size of those solved by Ryser's approach in the same time, despite the fact that Ryser's approach has exponential complexity even in the best case.

In this paper, we show that practically efficient algorithms for the
permanent can indeed be designed for large non-sparse matrices if the
matrix is represented compactly and manipulated efficiently using a
special class of data structures.  Specifically, we propose using
\emph{Algebraic Decision Diagrams}~\cite{BDGHMPS97} (ADDs) to
represent matrices, and design a version of Ryser's algorithm to work
on this symbolic representation of matrices.  This effectively gives
us a symbolic version of Ryser's algorithm, as opposed to existing
implementations that use an explicit representation of the matrix.
ADDs have been studied extensively in the context of formal
verification, and sophisticated libraries are available for compact
representation of ADDs and efficient implementation of ADD
operations~\cite{CUDD,van2017sylvan}.  The literature also contains
compelling evidence that reasoning based on ADDs and variants scales
to large instances of a diverse range of problems in practice,
cf.~\cite{BFGHMPF97,FMY97}. Our use of ADDs in Ryser's algorithm
leverages this progress for computing the permanent.  Significantly,
there are several sub-classes of matrices that admit compact
representations using ADDs, and our algorithm works well for all these
classes. Our empirical study provides evidence for the first time that
the frontier of practically efficient permanent computation can be
pushed well beyond the class of sparse matrices, to the classes of
dense matrices and, more generally, to matrices with ``similar''
rows. Coupled with a technique known as early abstraction, ADDs are
able to handle sparse instances as well. In summary, the symbolic
approach to permanent computation shows promise for both sparse and
dense classes of matrices, which are special cases of a notion of
row-similarity.

The rest of the paper is organized as follows: in Section 2 we
introduce ADDs and other concepts that we will use in this paper. We
discuss related work in Section 3 and present our algorithm and
analyze it in Section 4. Our empirical study is presented in Sections
5 and 6 and we conclude in Section 7.

%% file: Notation.tex
\section{Preliminaries}
\label{sec:prelim}
We denote by $ \mat{A} = (a_{ij}) $ an $ n\times n$ $0$-$1$ matrix,
which can also be interpreted as the bi-adjacency matrix of a
bipartite graph $ G_{\mat{A}} = (U \cup V, E)$ with an edge between
vertex $ i\in U $ and $ j\in V $ iff $ a_{ij} = 1$. We will denote the
$ i $th row of $ \mat{A} $ by $ r_i $. A perfect matching in $
G_{\mat{A}} $ is a subset $ \mathcal{M} \subseteq E $, such that for
all $ v\in (U\cup V) $, exactly one edge $ e \in \mathcal{M} $ is
incident on $ v $. We denote by $ \perm{\mat{A}} $ the permanent of $
\mat{A} $, and by $ \npm{G_{\mat{A}}} $, the number of perfect
matchings in $ G $. A well known fact is that $ \perm{\mat{A}} =
\npm{G_{\mat{A}}} $, and we will use these concepts interchangeably
when clear from context.

\subsection{Algebraic Decision Diagrams}
Let $X$ be a set of Boolean-valued variables.  An Algebraic Decision
Diagram (ADD) is a data structure used to compactly represent a
function of the form $f: 2^X\rightarrow\mathbb{R}$ as a Directed
Acyclic Graph (DAG).  ADDs were originally proposed as a
generalization of Binary Decision Diagrams (BDDs), which can only
represent functions of the form $g: 2^X \rightarrow \{0,1\}$.
Formally, an ADD is a $4$-tuple $(X,T,\pi,G)$ where $ X $ is a set of
Boolean variables, the finite set $ T \subset \mathbb{R}$ is called
the carrier set, $ \pi: X \rightarrow \mathbb{N} $ is the diagram
variable order, and $ G $ is a rooted directed acyclic graph
satisfying the following three properties:
\begin{enumerate}
	\item Every terminal node of $ G $ is labeled with an element of $ T $.
	\item Every non-terminal node of $ G $ is labeled with an element of $ X $ and has two outgoing edges labeled $0$ and $1$.
	\item For every path in $ G $, the labels of visited non-terminal nodes must occur in increasing order under $ \pi $.
\end{enumerate}
We use lower case letters $ f,g,\ldots $ to denote both functions from
Booleans to reals as well as the ADDs representing them. Many
operations on such functions can be performed in time polynomial in
the size of their ADDs. We list some such operations that will be used
in our discussion.
\begin{itemize}
	\item \emph{Product}: The product of two ADDs representing
          functions $f: 2^X\rightarrow\mathbb{R}$ and $g:
          2^Y\rightarrow\mathbb{R}$ is an ADD representing the
          function $f\cdot g: 2^{X\cup Y}\rightarrow\mathbb{R}$,
          where $ f\cdot g (\tau)$ is defined as $f(\tau \cap X) \cdot
          g(\tau \cap Y) $ for every $ \tau \in 2^{X\cup Y} $,
	\item \emph{Sum}: Defined in a way similar to the product.
	\item \emph{If-Then-Else (ITE):} This is a ternary operation
          that takes as inputs a BDD $ f $ and two ADDs $ g$ and $h $.
          $ITE(f,g,h) $ represents the function $f\cdot g + \neg f\cdot h$,
          and the corresponding ADD is obtained by substituting $ g
          $ for the leaf '1' of $ f $ and $ h $ for the leaf
          '0', and simplifying the resulting structure.
	\item \emph{Additive Quantification}: The existential
          quantification operation for Boolean-valued functions can be
          extended to real-valued functions by replacing disjunction
          with addition as follows. The additive quantification of $f:
          2^{X}\rightarrow\mathbb{R}$ is denoted as $ \exists x. f :
          2^{X\backslash\{x\}}\rightarrow\mathbb{R} $ and for $ \tau
          \in 2^{X\backslash\{x\}} $, we have $ \exists x. f(\tau) =
          f(\tau) + f(\tau \cup \{x\})$.
\end{itemize}
ADDs share many properties with BDDs. For example, there is a unique
minimal ADD for a given variable order $ \pi $, called the
\emph{canonical ADD}, and minimization can be performed in polynomial
time. Similar to BDDs, the variable order can significantly affect the
size of the ADD. Hence heuristics for finding good variable orders for
BDDs carry over to ADDs as well.  ADDs typically have lower
\emph{recombination efficiency}, i.e. number of shared nodes,
vis-a-vis BDDs.  Nevertheless, sharing or recombination of isomorphic
sub-graphs in an ADD is known to provide significant practical
advantages in representing matrices, vis-a-vis other competing data
structures.  The reader is referred to~\cite{BDGHMPS97} for a
nice introduction to ADDs and their applications.

\subsection{Ryser's Formula}
The permanent of $ \mat {A} $ can be calculated by the principle of
inclusion-exclusion using Ryser's formula:
\begin{math}
\perm{\mat{A}} = (-1)^n \sum_{S\subseteq[n]} (-1)^{|S|} \prod_{i=1}^{n} \sum_{j\in S} a_{ij}.
\end{math}
Algorithms implementing Ryser's formula on an explicit representation
of an arbitrary matrix $\mat{A}$ (not necessarily sparse) must
consider all $2^n$ subsets of $[n]$.  As a consequence, such
algorithms have at least exponential complexity.  Our experiments show
that even the best known existing algorithm implementing Ryser's formula for arbitrary matrices~\cite{nijenhuis2014combinatorial}, which iterates over the subsets of $[n]$ in Gray-code sequence, consistently times out after $1800$
seconds on a state-of-the-art computing platform when computing the
permanent of $n \times n$ matrices, with $n \ge 35$.

%
%
%
%

%% file: Related.tex
\section{Related Work}
\label{sec:rel_work}

Valiant showed that computing the permanent is
$\#P$-complete~\cite{Valiant79}. Subsequently, researchers
have considered restricted sub-classes of inputs in the quest for
efficient algorithms for computing the permanent, both from
theoretical and practical points of view.  We highlight some of
the important milestones achieved in this direction.

A seminal result is the Fisher-Temperly-Kastelyn algorithm~\cite{temperley1961dimer,kasteleyn1961statistics}, which computes the number of perfect matchings in planar graphs in PTIME. This result was subsequently extended to many other graph classes (c.f.~\cite{okamoto2009counting}). Following the work of Courcelle et al., a number of different width parameters have been proposed, culminating in the definition of ps-width~\cite{saether2015solving}, which is considered to be the most general notion of width~\cite{brault2014understanding}. Nevertheless, as with clique-width, it is not clear whether it lends itself to practically efficient algorithms. Bax and Franklin~\cite{bax2002permanent} gave a Las Vegas algorithm with better expected time complexity than Ryser's approach, but requiring $ \mathcal{O}(2^{n/2}) $ space.

For matrices with at most $ C\cdot n $ zeros, Servedio and Wan~\cite{servedio2005computing} presented a $ (2-\varepsilon)^n $-time and $ \mathcal{O}(n) $ space algorithm where $ \varepsilon $ depends on $ C $. Izumi and Wadayama~\cite{izumi2012new} gave an algorithm that runs in time $ \mathcal{O^*}(2^{(1-1/(\Delta \log \Delta))n}) $, where $ \Delta $ is the average degree of a vertex. On the practical side, in a series of papers, Liang, Bai and their co-authors~\cite{liang2004partially,liang2006hybrid,yue2013improved} developed  algorithms optimized for computing the permanent of the adjacency matrices of fullerenes, which are 3-regular graphs.

In recent years, practical techniques for propositional model counting
(\#SAT) have come of age. State-of-the-art exact model counters like
{\ds}~\cite{Muise2012} and {\dfr}~\cite{lagniez2017improved} also
incorporate techniques from knowledge compilation. A straightforward
reduction of the permanent to \#SAT uses a Boolean variable $x_{ij}$
for each $1$ in row $i$ and column $j$ of the input matrix $\mat{A}$,
and imposes Exact-One constraints on the variables in each row and
column.  This gives the formula $ F_{\perm{\mat{A}}} =
\bigwedge_{i\in[n]} ExactOne(\{x_{ij}: a_{ij}=1\}) \wedge\,\,
\bigwedge_{j\in[n]} ExactOne(\{x_{ij}: a_{ij}=1\}) $. Each solution to
$ F_{\perm{\mat{A}}} $ is a perfect matching in the underlying graph,
and so the number of solutions is exactly the permanent of the
matrix. A number of different encodings can be used for translating
Exact-One constraints to Conjunctive Normal Form (see Section
\ref{sec:algs}). We perform extensive comparisons
of our tool with {\dfr} and {\ds} with six such encodings.


%% file: Algorithm.tex
\section{Representing Ryser's Formula Symbolically}
As noted in Sec. \ref{sec:prelim}, an explicit implementation of Ryser's formula iterates over all $2^n$ subsets of columns and its complexity is in $\Theta(n\cdot 2^n)$.  Therefore, any such implementation takes exponential time even in the best case. A natural question to ask is whether we can do better through a careful selection of subsets over which to iterate. This principle was used for the case of sparse matrices by Servedio and Wan~\cite{servedio2005computing}. Their idea was to avoid those subsets for which the row-sum represented by the innermost summation in Ryser's formula, is zero for at least one row, since those terms do not contribute to the outer sum in Ryser's formula. Unfortunately, this approach does not help for non-sparse matrices, as very few subsets of columns (if any) will yield a zero row-sum. 

It is interesting to ask if we can exploit similarity of rows (instead
of sparsity) to our advantage. Consider the ideal case of an $n\times
n $ matrix with \emph{identical rows}, where each row has $k~(\le n)$
$1$s. For any given subset of columns, the row-sum is clearly the same
for all rows, and hence the product of all row-sums is simply the
$n^{th}$ power of the row-sum of one row. Furthermore, there are only
$k+1$ distinct values ($0$ through $k$) of the row-sum, depending on
which subset of columns is selected. The number of $r$-sized column
subsets that yield row-sum $j$ is clearly ${k \choose j}\cdot{n-k
  \choose r-j}$, for $0 \le j \le k$ and $j \le r \le n-k+j$.  Thus,
we can directly compute the permanent of the matrix via Ryser's
formula as
\begin{math}
\perm{\mat{A}} = (-1)^n \sum_{j=0}^{k}\sum_{r=j}^{n-k+j} (-1)^{r} {k \choose j}\cdot {n-k \choose r-j}\cdot j^n
\end{math}.
This equation has a more compact representation than the explicit
implementation of Ryser's formula, since the outer summation is over
$(k+1).(n-k+1)$ terms instead of $2^n$ terms.

Drawing motivation from the above example, we propose using
memoization to simplify the permanent computation of matrices with
similar rows.  Specifically, if we compute and store the row-sums for
a subset $S_1\subset[n]$ of columns, then we can potentially reuse
this information when computing the row-sums for subsets $S_2 \supset
S_1$.  We expect storage requirements to be low when the rows are
similar, as the partial sums over identical parts of the rows will have a compact representation, as shown above.

While we can attempt to hand-craft a concrete algorithm using this
idea, it turns out that ADDs fit the bill perfectly. We introduce
Boolean variables $x_j$ for each column $1\le j \le n $ in the
matrix. We can represent the summand $(-1)^{|S|} \prod_{i=1}^{n}
\sum_{j\in S} a_{ij} $ in Ryser's formula as a function $ f_{Ryser} : 2^X
\rightarrow \mathbb{R} $ where for a subset of columns $\tau\in 2^X
$, we have $ f_{Ryser}(\tau) = (-1)^{|\tau|} \prod_{i=1}^{n}
\sum_{j\in \tau} a_{ij} $. The outer sum in Ryser's formula is then
simply the Additive Quantification of $f_{Ryser}$ over all variables
in $ X $. The permanent can thus be denoted by the following equation:
\begin{equation}
\label{eqn:perm}
\perm{\mat{A}} = (-1)^n \,\,.\,\, \exists x_1,x_2,\ldots x_n. (f_{Ryser})
\end{equation}

\begin{figure}
	\centering
	\begin{subfigure}{0.32\textwidth}
		\includegraphics[width=\textwidth]{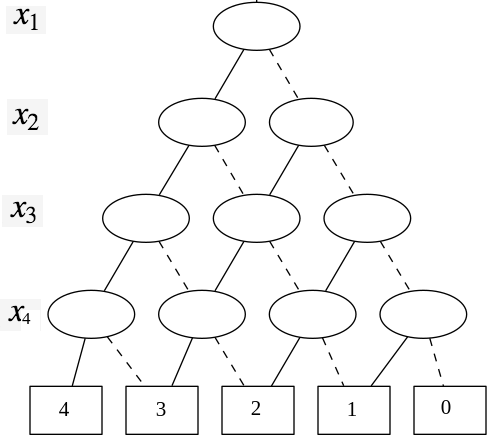}
		\caption{}
	\end{subfigure}
	\begin{subfigure}{0.32\textwidth}
		\includegraphics[width=\textwidth]{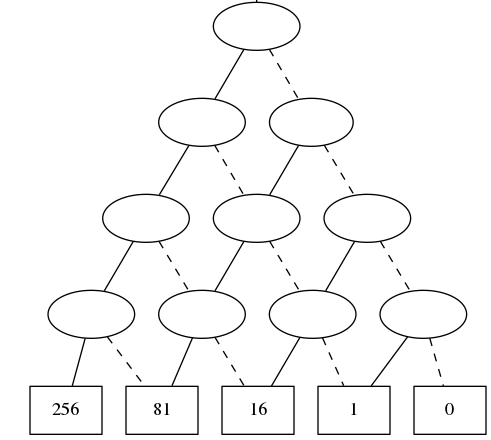}
		\caption{}
	\end{subfigure}
	\begin{subfigure}{0.32\textwidth}
		\includegraphics[width=\textwidth]{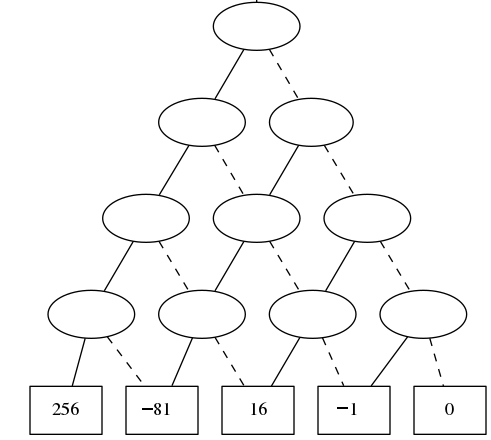}
		\caption{}
	\end{subfigure}
	\caption{(a) $ f_{RS} $, (b) $ f_{RSP} $ and (c) $ f_{Ryser} $ for a $ 4 \times 4 $ matrix of all 1s}
	\label{fig:add}
	\vspace{-0.5cm}
\end{figure}

We can construct an ADD for $ f_{Ryser} $ incrementally as follows:
\begin{itemize}
\item \textbf{Step 1}: For each row $ r_i $ in the matrix, construct the Row-Sum ADD $ f_{RS}^{r_i} $ such that $ f_{RS}^{r_i}(\tau) = \sum_{j: a_{ij} = 1} \mathbbm{1}_{\tau} (x_j) $, where $  \mathbbm{1}_{\tau} (x_j) $ is the indicator function taking the value 1 if $ x_j \in \tau $, and zero otherwise. This ADD can be constructed by using the sum operation on the variables $ x_j $ corresponding to the $1$ entries in row $ r_i $.
	\item \textbf{Step 2}: Construct the Row-Sum-Product ADD $ f_{RSP} = \prod_{i = 1 }^{n} f_{RS}^{r_i} $ by applying the product operation on all the Row-Sum ADDs
	\item \textbf{Step 3}: Construct the Parity ADD $ f_{PAR} = ITE(\bigoplus_{j=1}^{n} x_j, -1, +1) $, where $\bigoplus$ represents exclusive-or.  This ADD represents the $ (-1)^{|S|} $ term in Ryser's formula.
	\item \textbf{Step 4}: Construct $ f_{Ryser} = f_{RSP} . f_{PAR} $ using the product operation.
\end{itemize}

Finally, we can additively quantify out all variables in $ f_{Ryser} $ and multiply the result by $ (-1)^n $ to get the permanent, as given by Equation~\ref{eqn:perm}.

The size of the ADD $ f_{RSP} $ will be the smallest when the ADDs $
f_{RS}^{r_i} $ are exactly the same for all rows $r_i$, i.e. when all
rows of the matrix are identical. In this case, the ADDs $f_{RS}^{r_i}
$ and $ f_{RSP} $ will be isomorphic; the values at the leaves of
$f_{RSP}$ will simply be the $n^{th}$ power of the values at the
corresponding leaves of $f_{RS}^{r_i}$.  An example illustrating this
for a $4\times 4$ matrix of all $1$s is shown in
Fig. \ref{fig:add}. Each level of the ADDs in this figure corresponds
to a variable (shown on the left) for a column of the matrix. A solid
edge represents the 'true' branch while a dotted edge represents the
'false' branch. Observe that sharing of isomorphic subgraphs allows
each of these ADDs to have $10$ internal nodes and $5$ leaves, as
opposed to $15$ internal nodes and $16$ leaves that would be needed
for a complete binary tree based representation.

The ADD representation is thus expected to be compact when the
rows are ``similar''.  Dense matrices can be thought of as a special
case: starting with a matrix of all $1$s (which clearly has all rows
identical), we change a few $1$s to $0$s. The same idea can be applied
to sparse matrices as well: starting with a matrix of all $0$s (once
again, identical rows), we change a few $0$s to $1$s. The case of very
sparse matrices is not interesting, however, as the permanent (or
equivalently, count of perfect matchings in the corresponding
bipartite graph) is small and can be computed by naive enumeration.
Interestingly, our experiments show that as we reduce the sparsity of
the input matrix, constructing $f_{RSP}$ and $f_{Ryser}$ in a
monolithic fashion as discussed above fails to scale, since the sizes
of ADDs increase very sharply. Therefore we need additional
machinery.

First, we rewrite Equation~\ref{eqn:perm} in terms of the intermediate ADDs as:
\begin{equation}
\label{eqn:perm2}
\perm{\mat{A}} = (-1)^n \,\,.\,\, \exists x_1,x_2,\ldots x_n. \left(f_{PAR}\cdot\prod_{i=1}^{n} f_{RS}^{r_i}\right)
\end{equation}

We then employ the principle of early abstraction to compute
$f_{Ryser}$ incrementally.  Note that early abstraction has been used
successfully in the past in the context of SAT solving~\cite{PV04a},
and recently for weighted model counting using ADDs in a technique called {\addmc}~\cite{addmc}. The formal statement of the principle of early abstraction is given in the
following theorem.

\begin{theorem}{\rm ~\cite{addmc}}
	Let $ X $ and $ Y $ be sets of variables and $ f:2^X\rightarrow\mathbb{R} $, $ g:2^Y\rightarrow\mathbb{R} $. For all $ x\in X\setminus Y $, we have 
	\begin{math}
	\exists_x (f\cdot g) = (\exists_x (f))\cdot g
	\end{math}
\end{theorem}

Since the product operator is associative and additive quantification
is commutative, we can rearrange the terms of Equation~\ref{eqn:perm2}
in order to apply early abstraction. This idea is implemented in
Algorithm $\radd$, which is motivated by the weighted model counting
algorithm in~\cite{addmc}.

\begin{algorithm}
	\caption{$\radd(\mat{A},\pi,\cvo)$}
	\label{alg:radd}
	\begin{algorithmic}[1]
		\State $ m \gets max_{x\in X} \,\,\cvo(x) $;
		\For {$i = m, m-1, \ldots , 1$}
			\State $ \kappa_i \gets \{f_{RS}^{r} : r$ is a row in $\mat{A}$ and $\clusterrank(r,\cvo)=i \} $;\label{lin:part}
		\EndFor
		\State $ f_{Ryser} \gets f_{PAR} $;\label{lin:par}
		\Comment $ f_{PAR} $ and each $ f_{RS}^{r} $ are constructed using the diagram variable order $\pi $
		\For {$ i = 1,2,\ldots,m $}
			\If {$ \kappa_i \ne \emptyset $}
				\For {$ g \in \kappa_i $} \label{lin:loop}
					\State $ f_{Ryser} \gets f_{Ryser} \cdot g $;
				\EndFor	
				\For {$ x \in Vars(f_{Ryser}) $} \label{lin:early}
					\If {$ x \not \in (Vars(\kappa_{i+1}) \cup \ldots \cup Vars(\kappa_{m})) $}
						\State $ f_{Ryser} \gets \exists_x(f_{Ryser}) $
					\EndIf
				\EndFor
			\EndIf
		\EndFor
		\State \Return $ (-1)^n\times f_{Ryser}(\emptyset) $ \label{lin:ret}
	\end{algorithmic}
\end{algorithm}

Algorithm $\radd$ takes as input a 0-1 matrix $ \mat{A} $, a diagram
variable order $ \pi $ and a cluster rank-order $ \cvo $. $ \cvo $ is an ordering of variables which is used to heuristically partition rows of $\mat{A}$ into clusters using a function $\clusterrank$, where all rows in a cluster get the same rank.  Intuitively, rows that
are almost identical are placed in the same cluster, while those that
differ significantly are placed in different clusters.  Furthermore,
the clusters are ordered such that there are non-zero columns in
cluster $i$ that are absent in the set of non-zero columns in clusters
with rank $> i$.  As we will soon see, this facilitates keeping the
sizes of ADDs under control by applying early abstraction.

Algorithm $\radd$ proceeds by first partitioning the Row-Sum ADDs of
the rows $ \mat{A} $ into clusters according to their cluster rank in
line \ref{lin:part}. Each Row-Sum ADD is constructed according to the
diagram variable order $ \pi $. The ADD $ f_{Ryser} $ is constructed
incrementally, starting with the Parity ADD in line \ref{lin:par}, and
multiplying the Row-Sum ADDs in each cluster $ \kappa_i $ in the loop
at line \ref{lin:loop}. However, unlike the monolithic approach, early
abstraction is carried out within the loop at line
\ref{lin:early}. Finally, when the execution reaches
line~\ref{lin:ret}, all variables representing columns of the input
matrix have been abstracted out.  Therefore, $ f_{Ryser} $ is an ADD
with a single leaf node that contains the (possibly negative) value of
the permanent. Following Equation~\ref{eqn:perm2}, the algorithm
returns the product of $ (-1)^n $ and $f_{Rsyer}(\emptyset)$.

The choice of the function $ \clusterrank $ and the cluster rank-order $ \cvo $ significantly affect the performance of the algorithm.
A number of heuristics for determining $\clusterrank$ and $\cvo$ have
been proposed in literature, such as Bucket Elimination~\cite{dechter1999bucket}, and Bouquet's
Method~\cite{bouquet1999gestion} for cluster ranking, and MCS~\cite{tarjan1984simple}, LexP~\cite{koster2001treewidth} and LexM~\cite{koster2001treewidth} for variable
ordering. Further details and a rigorous comparison of these
heuristics are presented in~\cite{addmc}. Note that if we assign the
same cluster rank to all rows of the input matrix, Algorithm $\radd$
reduces to one that constructs all ADDs monolithically, and does not
benefit from early abstraction.

\subsection{Implementation Details}
\label{sec:radd_impl}
We implemented Algorithm \ref{alg:radd} using the library
Sylvan~\cite{van2017sylvan} since unlike CUDD~\cite{CUDD}, Sylvan
supports arbitrary precision arithmetic -- an essential feature to
avoid overflows when the permanent has a large value. Sylvan supports
parallelization of ADD operations in a multi-core environment.  In
order to leverage this capability, we created a parallel version of
$\radd$ that differs from the sequential version only in that it uses
the parallel implementation of ADD operations natively provided by
Sylvan.  Note that this doesn't require any change to Algorithm
$\radd$, except in the call to Sylvan functions. While other
non-ADD-based approaches to computing the permanent can be
parallelized as well, we emphasize that it is a non-trivial task in
general, unlike using Sylvan. We refer to our sequential and parallel
implementations for permanent computation as {\radd} and {\raddp}
respectively, in the remainder of the discussion. We implemented our
algorithm in C++, compiled under GCC v6.4 with the O3 flag. We
measured the wall-times for both algorithms. Sylvan also supports arbitrary precision floating
point computation, which makes it easy to extend {\radd} for computing
permanent of real-valued matrices.  However, we leave a detailed
investigation of this for future work.

%% file: Experiments.tex
\section{Experimental Methodology}

The objective of our empirical study was to evaluate {\radd} and
{\raddp} on randomly generated instances (as done
in~\cite{liang2006hybrid}) and publicly available structured instances
(as done in~\cite{liang2004partially,yue2013improved}) of $0$-$1$
matrices.

\subsection{Algorithm Suite}
\label{sec:algs}
As noted in Section \ref{sec:rel_work}, a number of different
algorithms have been reported in the literature for computing the
permanent of sparse matrices. Given resource constraints, it is
infeasible to include all of these in our experimental
comparisons. This is further complicated by the fact that many of
these algorithms appear not to have been implemented (eg:
\cite{servedio2005computing,izumi2012new}), or the code has not been
made publicly accessible (eg:
~\cite{liang2004partially,yue2013improved}). A fair comparison would
require careful consideration of several parameters like usage of
libraries, language of implementation, suitability of hardware etc. We had to arrive at an informed choice of algorithms, which we list below along with our rationale:
\begin{itemize}
	\item {\radd} and {\raddp}: For the dense and similar rows cases, we use the monolithic approach as it is sufficient to demonstrate the scalability of our ADD-based approach. For sparse instances, we employ Bouquet's Method (List)~\cite{bouquet1999gestion} clustering heuristic along with MCS cluster rank-order~\cite{tarjan1984simple} and we keep the diagram variable order the same as the indices of columns in the input matrix (see~\cite{addmc} for details about the heuristics). We arrived at these choices through preliminary experiments. We leave a detailed comparison of all combinations for future work.
	\item \textit{Explicit Ryser's Algorithm}: We implemented Nijenhuis and Wilf's version~\cite{nijenhuis2014combinatorial} of Ryser's formula using Algorithm H from {\cite{knuth2004generating}} for generating the Gray code sequence. Our implementation, running on a state-of-the-art computing platform (see Section~\ref{sec:setup}), is able to compute the permanent of all matrices with $ n \le 25 $ in under 5 seconds. For $ n = 30 $, the time shoots up to approximately $460$ seconds and for $n \ge 34$, the time taken exceeds 1800 seconds (time out for our experiments).  Since the performance of explicit Ryser's algorithm depends only on the size of the matrix, and is unaffected by its structure, sparsity or row-similarity, this represents a complete characterization of the performance of the explicit Ryser's algorithm.  Hence, we do not include it in our plots.
	\item \textit{Propositional Model Counters}: Model counters that employ techniques from SAT-solving as well as knowledge compilation, have been shown to scale extremely well on large CNF formulas from diverse domains. Years of careful engineering have resulted in counters that can often outperform domain-specific approaches. We used two state-of-the-art exact model counters, viz. {\dfr}~\cite{lagniez2017improved} and {\ds}~\cite{Muise2012}, for our experiments. We experimented with 6 different encodings for At-Most-One constraints: (1) Pairwise~\cite{Biere2009Handbook}, (2) Bitwise~\cite{Biere2009Handbook}, (3) Sequential Counter~\cite{sinz2005towards}, (4) Ladder~\cite{gent2004new,ansotegui2004mapping}, (5) Modulo Totalizer~\cite{ogawa2013modulo} and (6) Iterative Totalizer~\cite{martins2014incremental}. We also experimented with {\addmc}, an ADD-based model counter~\cite{addmc}.  However, it failed to scale beyond matrices of size $25$; ergo we do not include it in our study. 
\end{itemize}
We were unable to include the parallel \#SAT counter
countAtom~\cite{burchard2015laissez} in our experiments, owing to
difficulties in setting it up on our compute set-up. However, we could
run countAtom on a slightly different set-up with 8 cores instead of
12, and 16GB memory instead of 48 on a few sampled dense and
similar-row matrix instances.  Our experiments showed that
countAtom timed out on all these cases. We leave a more thorough and
scientific comparison with countAtom for future work.

\subsection{Experimental Setup}\label{sec:setup}
Each experiment (sequential or parallel) had exclusive access to a
Westemere node with 12 processor cores running at 2.83 GHz with 48 GB
of RAM. We capped memory usage at 42 GB for all tools. We implemented
explicit Ryser's algorithm in C++, compiled with GCC v6.4 with O3
flag. The {\radd} and {\raddp} algorithms were implemented as in
Section~\ref{sec:radd_impl}. {\raddp} had access to all $12$ cores for
parallel computation. We used the python library
PySAT~\cite{imms-sat18} for encoding matrices into CNF. We set the
timeout to $1800$ seconds for all our experiments.  For purposes of
reporting, we treat a memory out as equivalent to a time out.

\begin{table}
	\vspace{-0.5cm}
	\caption{Parameters used for generating random matrices}
	\resizebox{\textwidth}{!}{%
		\begin{tabular}{|l|c|c|c|c|c|}
			\hline
			\multicolumn{1}{|c|}{\textbf{Experiment}} & \textbf{\begin{tabular}[c]{@{}c@{}}Matrix Size\\ n\end{tabular}} & \textbf{$C_f$, where $C_f\cdot n$ matrix entries flipped}          & \textbf{\begin{tabular}[c]{@{}c@{}}Starting Matrix\\ Row Density $ \rho $\end{tabular}} & \textbf{\#Instances} & \textbf{\begin{tabular}[c]{@{}c@{}}Total\\ Benchmarks\end{tabular}} \\ \hline
			Dense                                     & 30, 40, 50, 60, 70                                               & 1, 1.1, 1.2, 1.3, 1.4   & 1                                                                              & 20                   & 500                                                                 \\ \hline
			Sparse                                    & 30, 40, 50, 60, 70                                               & 3.9, 4.3, 4.7, 5.1, 5.5 & 0                                                                              & 20                   & 500                                                                 \\ \hline
			Similar                                   & 40, 50, 60, 70, 80                                               & 1, 1.1, 1.2, 1.3, 1.4   & 0.7, 0.8, 0.9                                                                  & 15                   & 1125                                                                \\ \hline
		\end{tabular}%
	}
	\label{tab:params}
	\vspace{-0.5cm}
\end{table}
\subsection{Benchmarks}
The parameters used for generating random instances are summarized in Table \ref{tab:params}.  We do not include matrices with $ n < 30 $ since the explicit Ryser's algorithm suffices (and often performs the best) for such matrices. The upper bound for $ n $ was chosen such that the algorithms in our suite either timed out or came close to timing out. For each combination of parameters, random matrix instances were sampled as follows:
\begin{enumerate}
\item We started with an $ n \times n $ matrix, where the first row
  had $ \rho\cdot n $ $1$s at randomly chosen column positions, and
  all other rows were copies of the first row.
	\item $ C_f\cdot n $ randomly chosen entries in the starting matrix are flipped i.e. $0$ flipped to $1$ and vice versa.
\end{enumerate}
For the dense case, we start with a matrix of all $1$s while for the
sparse case, we start with a matrix of all $0$s, and used intermediate
row density values for the similar-rows case. We chose higher values
for $ C_f $ in the sparse case because for low values, the bipartite
graph corresponding to the generated matrix had very few perfect
matchings (if any), and these could be simply counted by
enumeration. We generated a total of $2125$ benchmarks covering a
broad range of parameters. For all generated instances, we ensured
that there was at least one perfect matching, since the case with zero
perfect matchings can be easily solved in polynomial time by
algorithms like Hopcroft-Karp~\cite{hopcroft1973n}. In order to avoid
spending inordinately large time on failed experiments, if an
algorithm timed out on all generated random instances of a particular
size, we also report a time out for that algorithm on all larger
instances of that class of matrices. We also double-check this by
conducting experiments with the same algorithm on a few randomly
chosen larger instances.

The SuiteSparse Matrix Collection~\cite{davis2011university} is a well
known repository of structured sparse matrices that arise from
practical applications. We found $26$ graphs in this suite with vertex
count between $30$ and $100$, of which $18$ had at least one perfect
matching. Note that these graphs are not necessarily bipartite;
however, their adjacency matrices can be used as benchmarks for
computing the permanent. A similar approach was employed
in~\cite{wang2012load} as well.

Fullerenes are carbon molecules whose adjacency matrices have been
used extensively by Liang et
al.~\cite{liang2004partially,wang2012load,yue2013improved} for
comparing tools for the permanent. We were able to find the
adjacency matrices of $ C_{60} $ and $ C_{100} $, and have used
these in our experiments.
  
\section{Results}
We first study the variation of running time of {\radd} with the size
of ADDs involved. Then we compare the running times of various
algorithms on sparse, dense and similar-row matrices, as well as on
instances from SuiteSparse Matrix Collection and on adjacency matrices
of fullerenes $ C_{60} $ and $ C_{100} $.  The total computational
effort of our experiments exceeds $2500$ hours of wall clock time on
dedicated compute nodes.

\subsection{ADD size vs time taken by $\radd$}
\begin{figure}[t]
	\centering
	\begin{subfigure}{0.32\textwidth}
		\includegraphics[width=\textwidth]{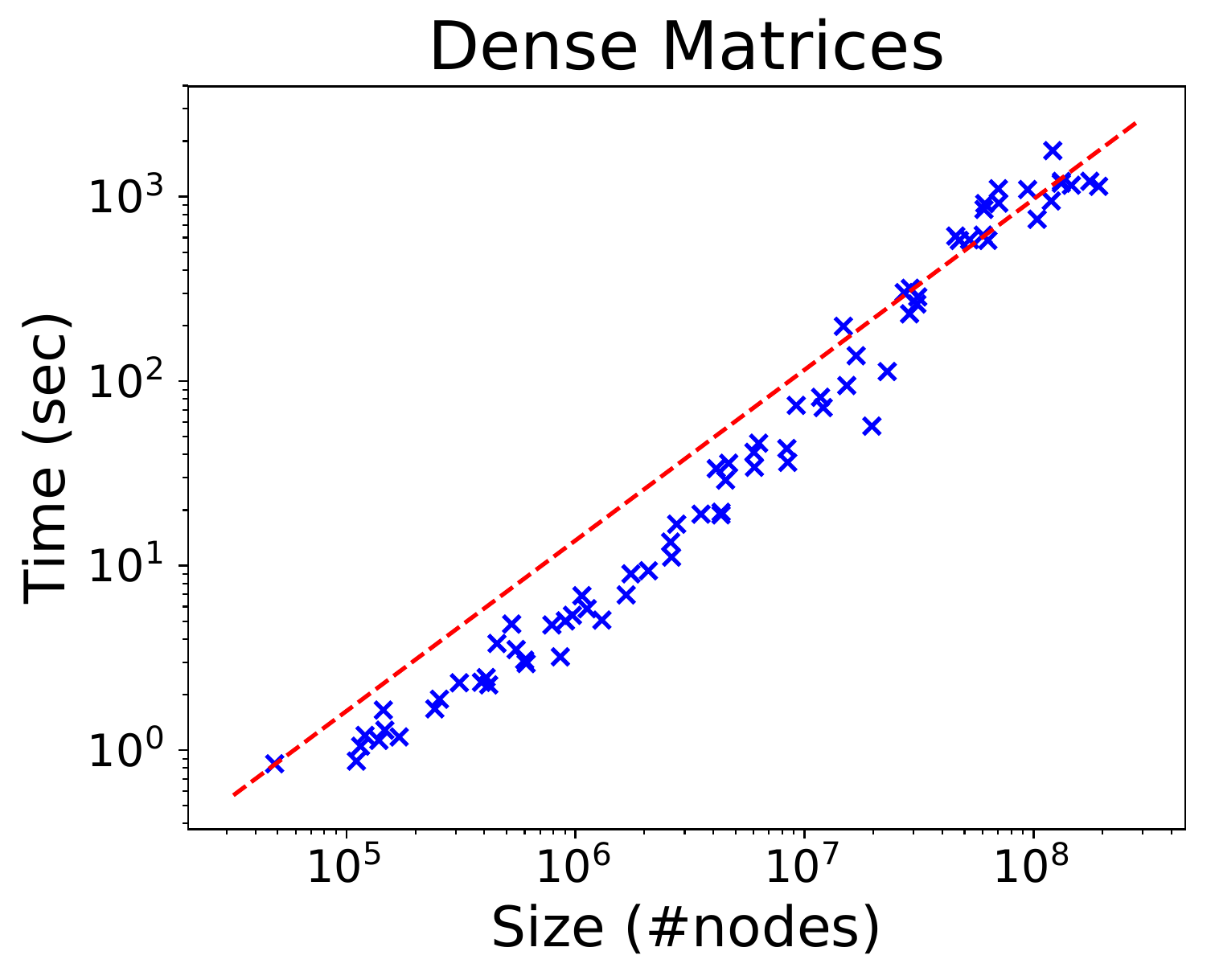}
	\end{subfigure}
	\begin{subfigure}{0.32\textwidth}
		
		\includegraphics[width=\textwidth]{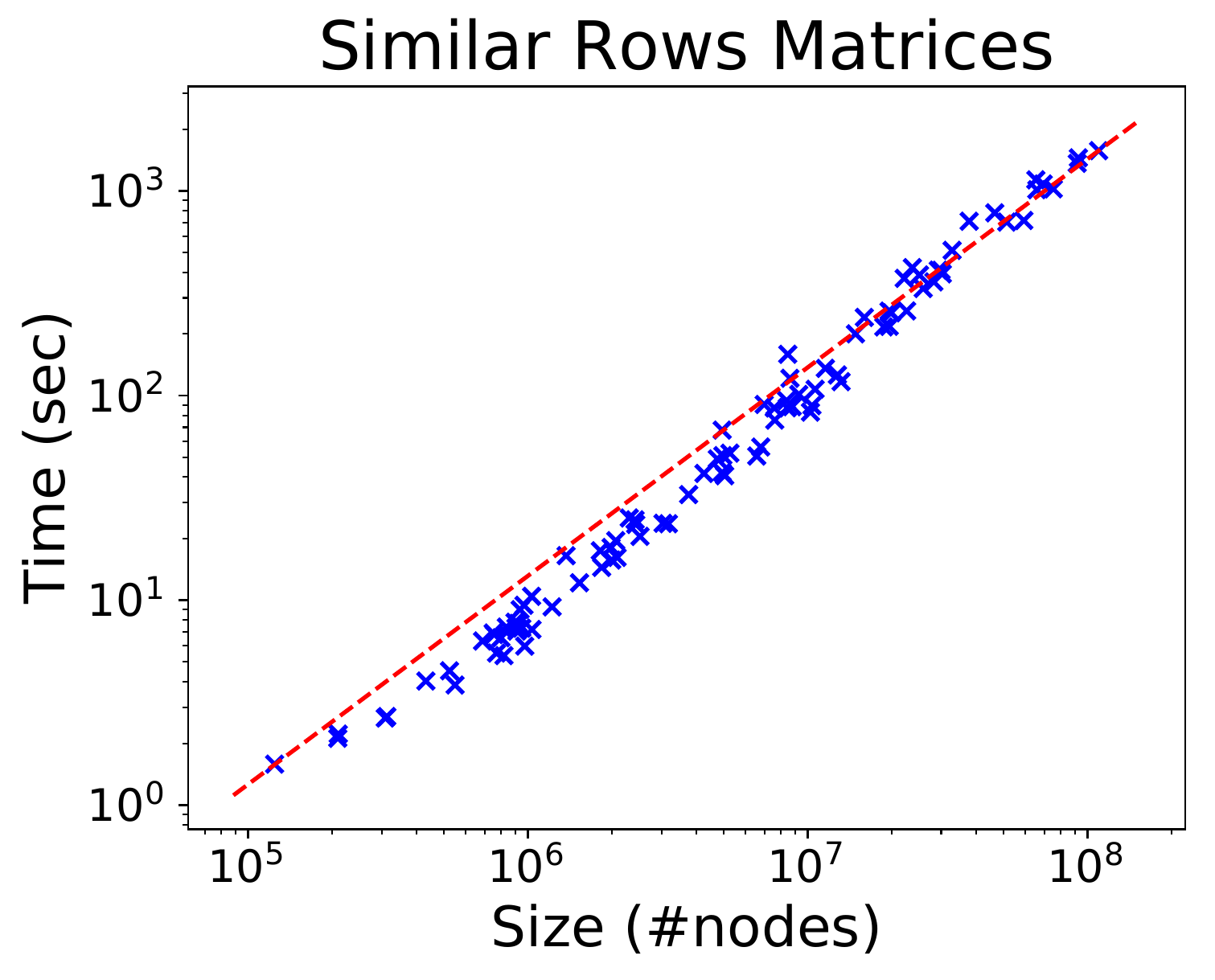}
	\end{subfigure}
	\begin{subfigure}{0.32\textwidth}
		\includegraphics[width=\textwidth]{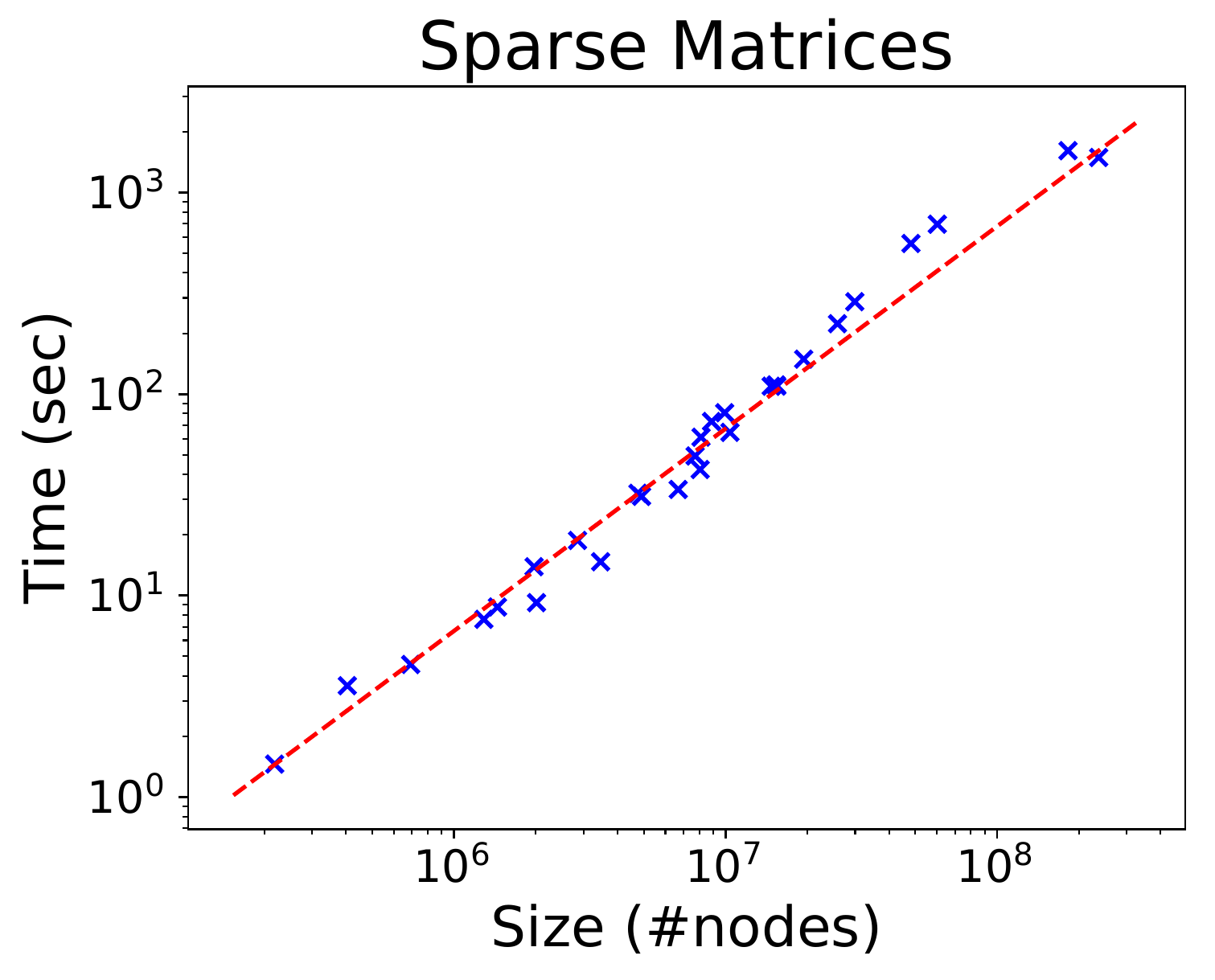}
	\end{subfigure}
	\caption{Comparison of ADD Size vs. Time taken for a subset of random benchmarks}
	\label{fig:scatter}
	\vspace{-0.2cm}
\end{figure}
In order to validate the hypothesis that the size of the ADD
representation is a crucial determining factor of the performance of
$\radd$, we present $3$ scatter-plots (Fig. \ref{fig:scatter}) for a
subset of $100$ instances, of each of the dense, sparse and
similar-rows cases. In each case, the $100$ instances cover the entire
range of $ C_f $ and $ n $ used in Table \ref{tab:params}, and we plot
times only for instances that didn't time out.  The plots show that
there is very strong correlation between the number of nodes in the
ADDs and the time taken for computing the permanent, supporting our
hypothesis.

\begin{figure}
	\centering
	\begin{subfigure}{0.48\textwidth}
		\includegraphics[width=\textwidth]{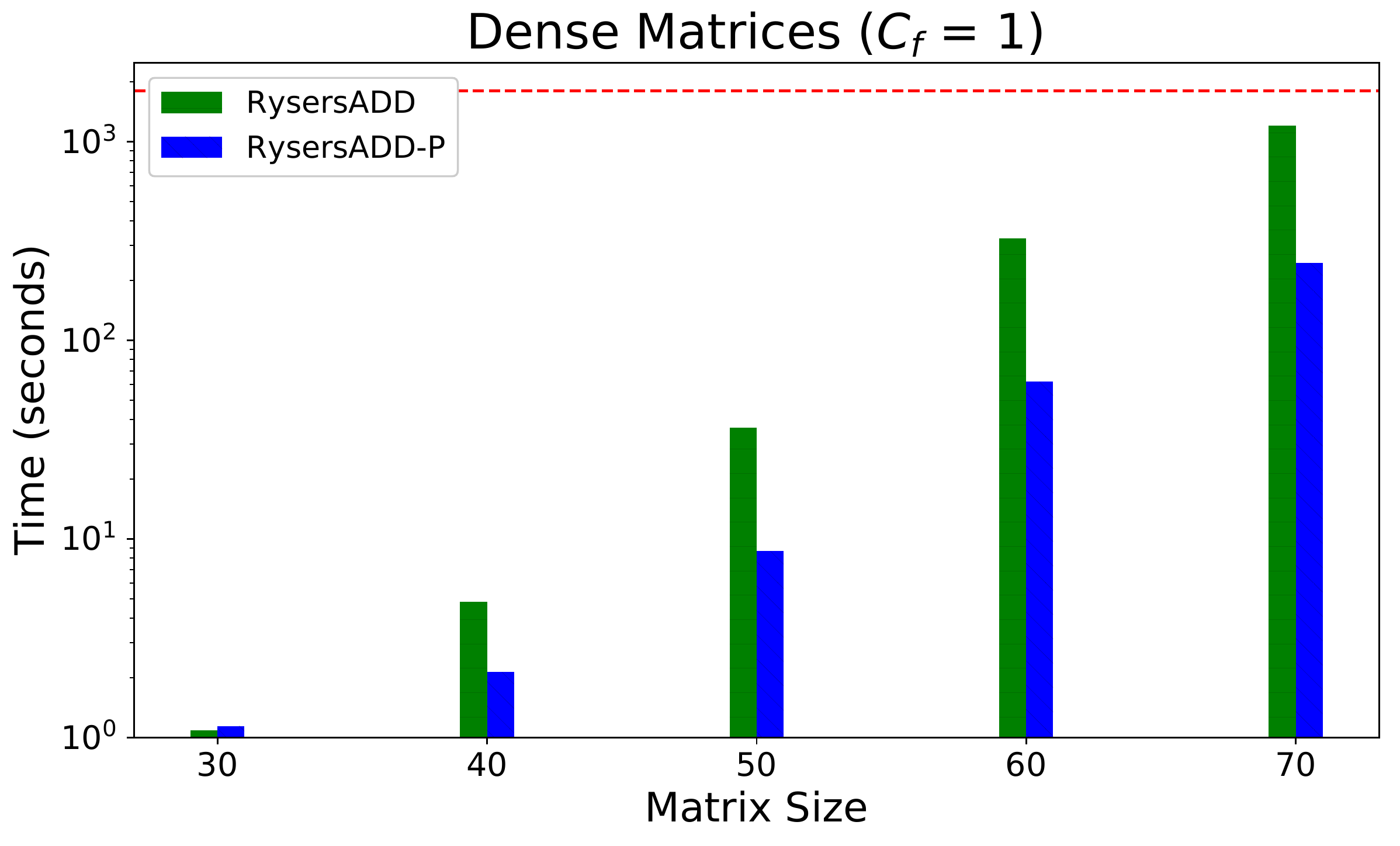}
	\end{subfigure}
	\hfill
	\begin{subfigure}{0.48\textwidth}
		
		\includegraphics[width=\textwidth]{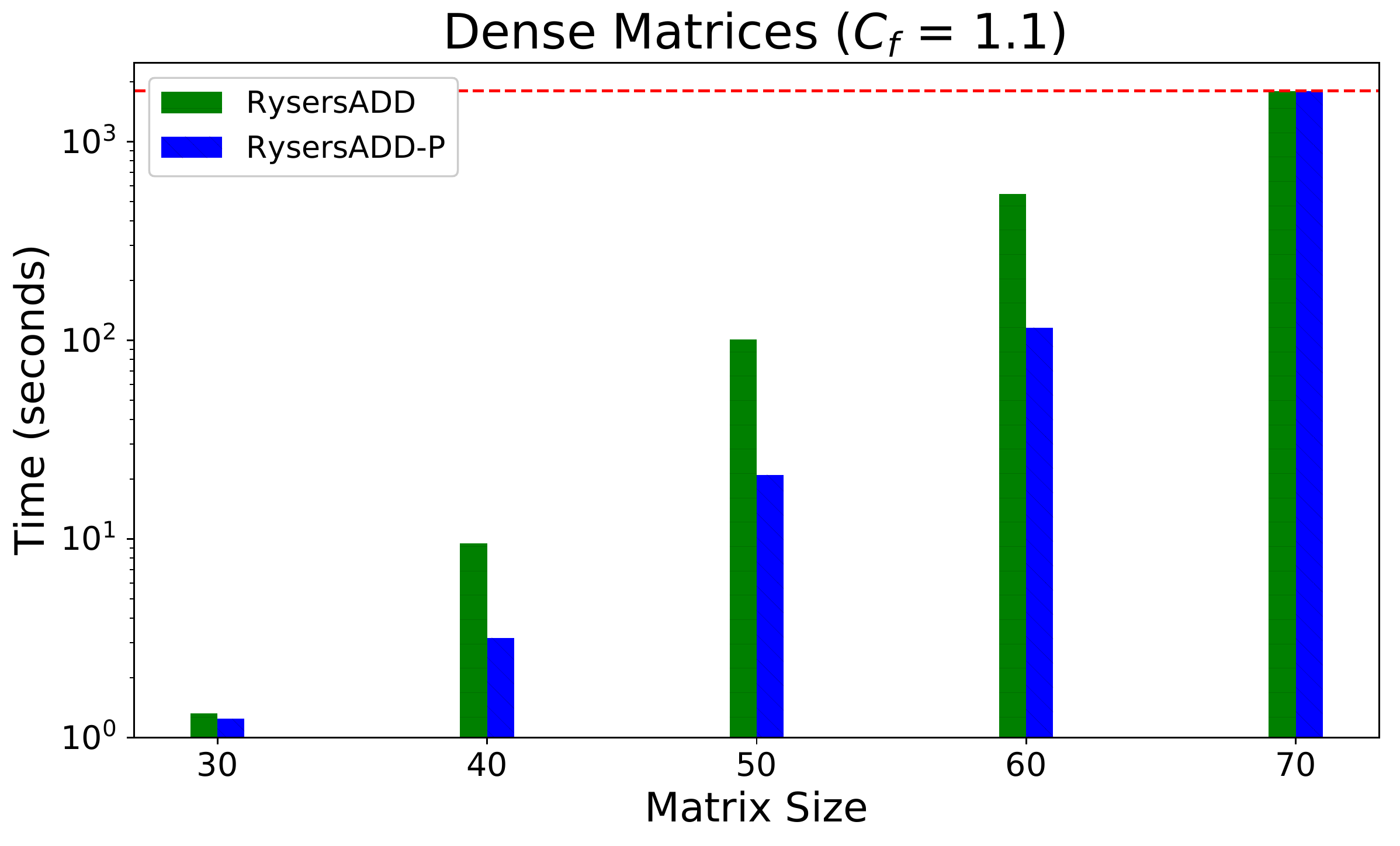}
	\end{subfigure}
	\begin{subfigure}{0.48\textwidth}
		\includegraphics[width=\textwidth]{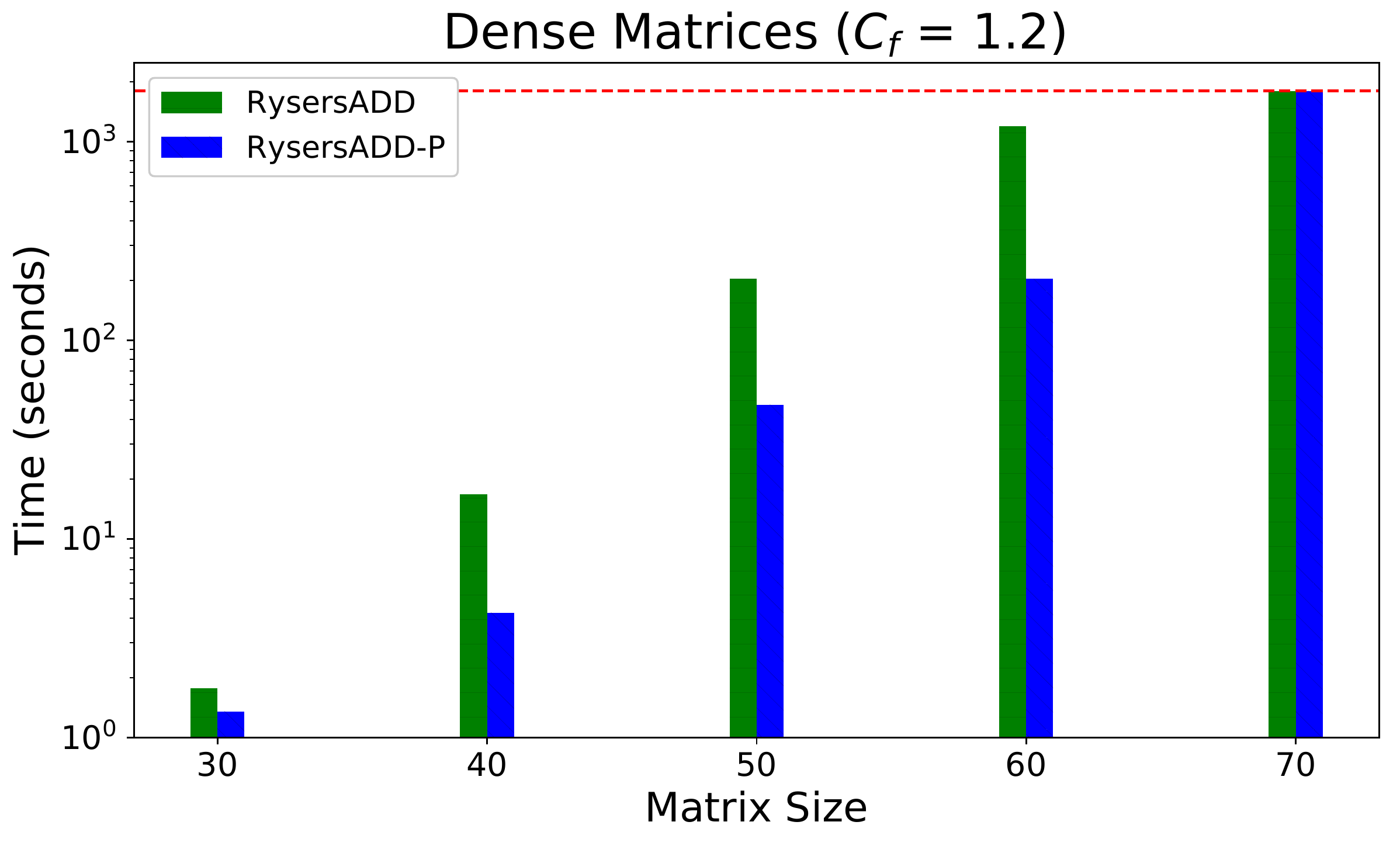}
	\end{subfigure}
	\hfill
	\begin{subfigure}{0.48\textwidth}
		
		\includegraphics[width=\textwidth]{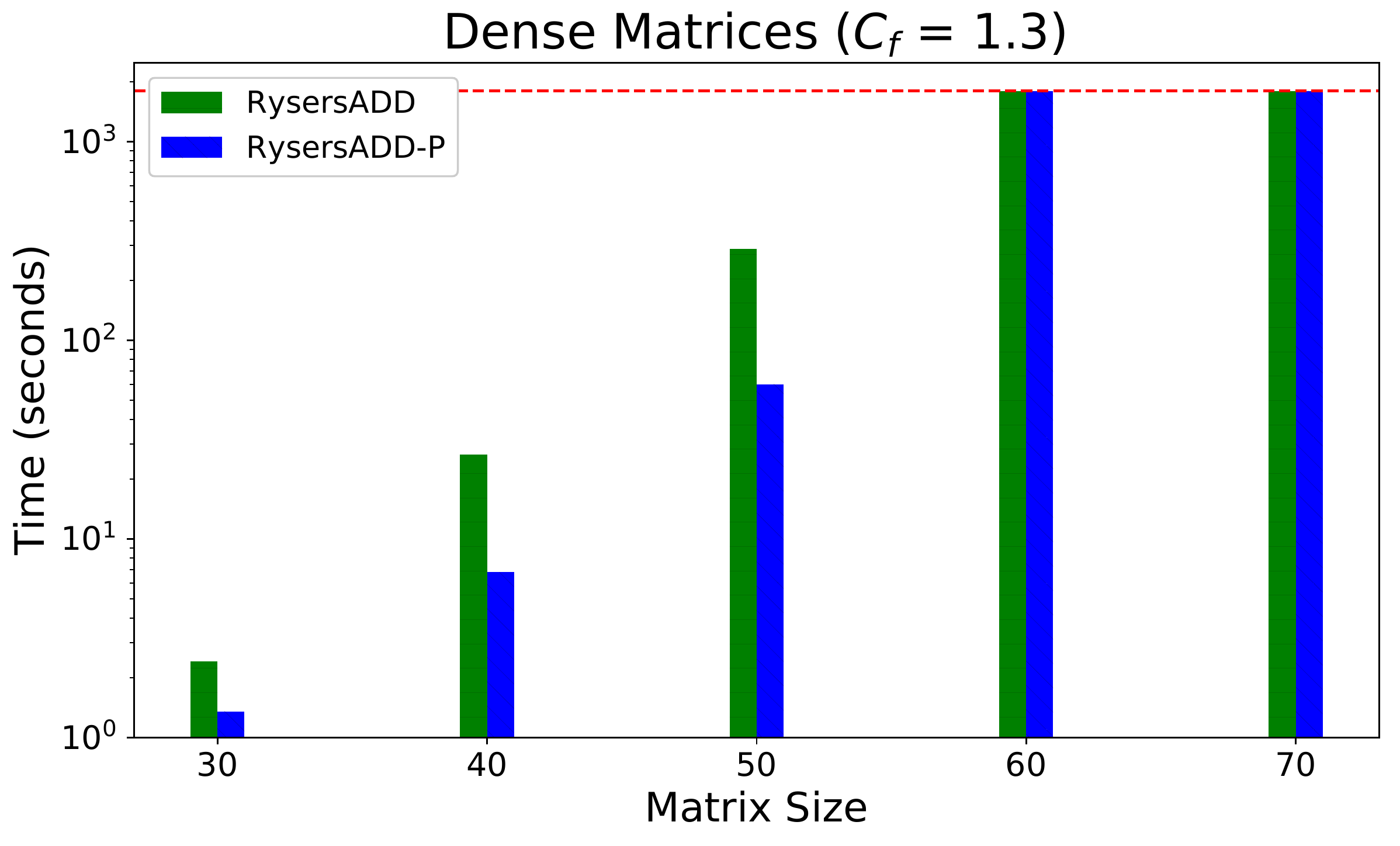}
	\end{subfigure}
	\caption{Performance on Dense Matrices. {\dfr}, {\ds} (not shown) timeout on all instances}
	\label{fig:dense}
	\vspace{-1cm}
\end{figure}  
\subsection{Performance on dense matrices}
We plot the median running time of $\radd$ and $\raddp$ against the
matrix size $n$ for dense matrices with $ C_f \in \{1,1.1,1.2,1.3\} $ in
Fig. \ref{fig:dense}.  We only show the running times of {\radd} and
{\raddp}, since {\dfr} and {\ds} were unable to solve any instance of
size $30$ for all $6$ encodings. We observe that the running time of
both the ADD-based algorithms increases with $ C_f $. This trend
continues for $ C_f = 1.4 $, which we omit for lack of space. {\raddp}
is noticeably faster than {\radd}, indicating that the native
parallelism provided by Sylvan is indeed effective.

\begin{figure}[t]
	\centering
	\begin{subfigure}{0.48\textwidth}
		\includegraphics[width=\textwidth]{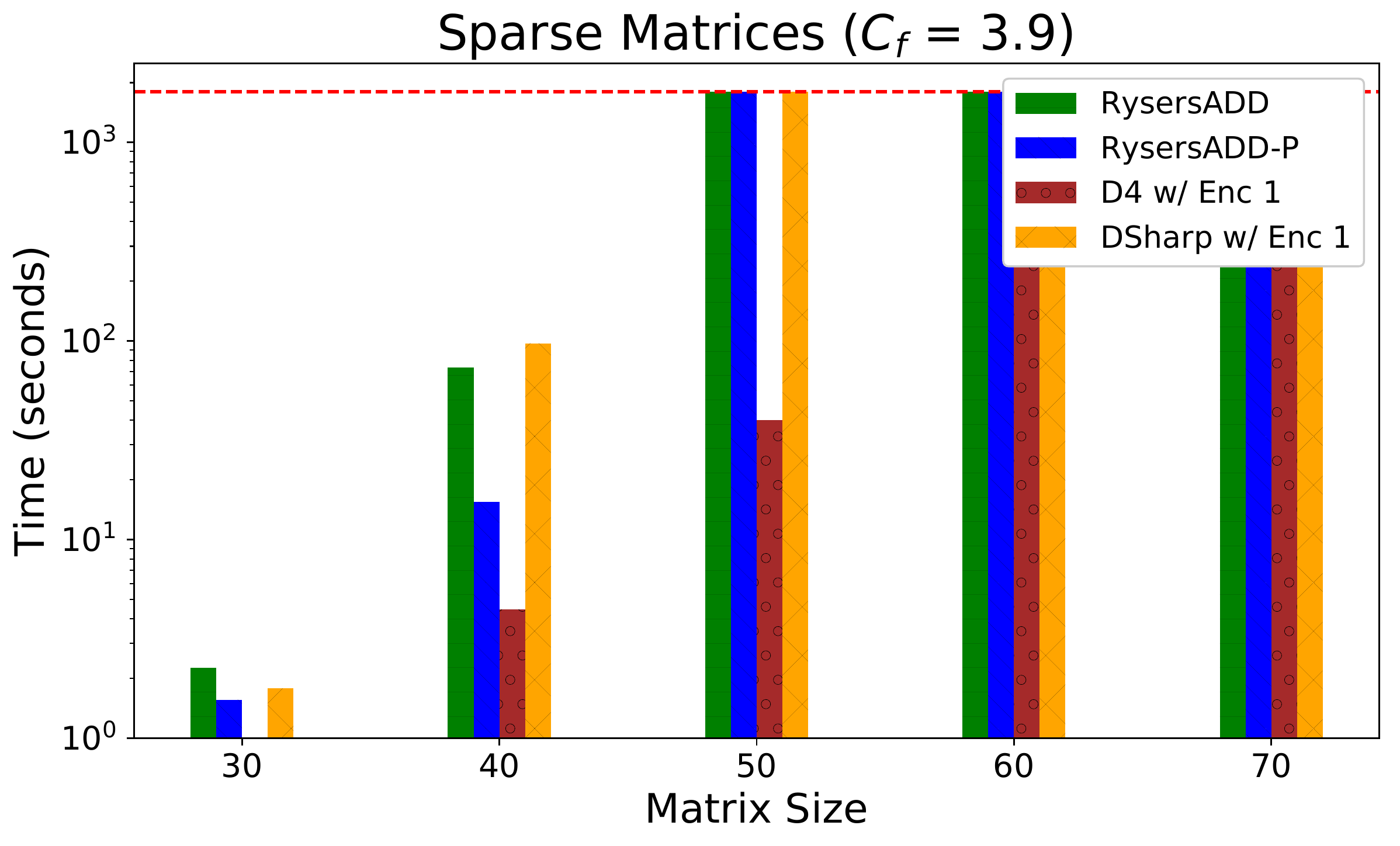}
	\end{subfigure}
	\hfill
	\begin{subfigure}{0.48\textwidth}
		
		\includegraphics[width=\textwidth]{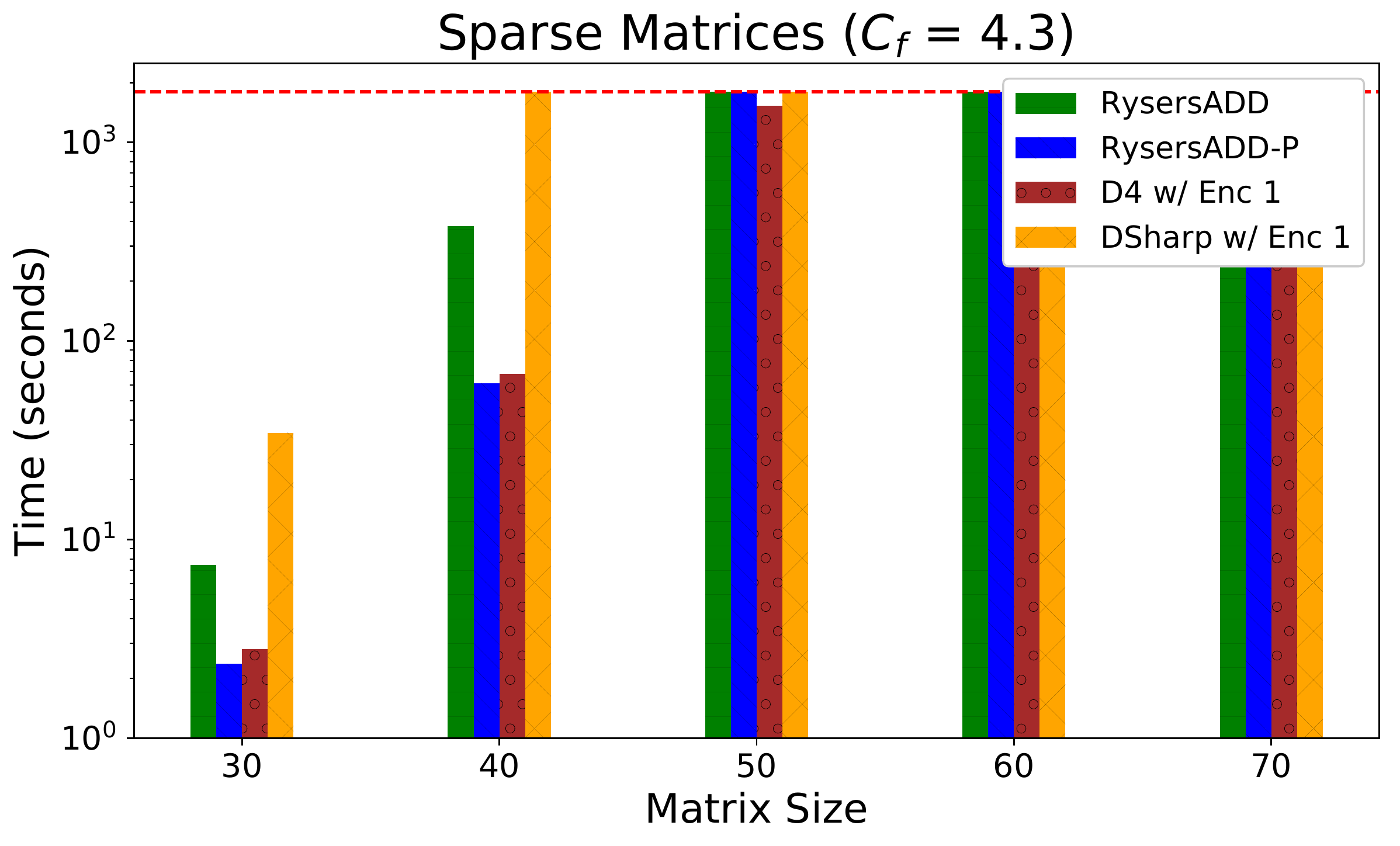}
	\end{subfigure}
	\begin{subfigure}{0.48\textwidth}
		\includegraphics[width=\textwidth]{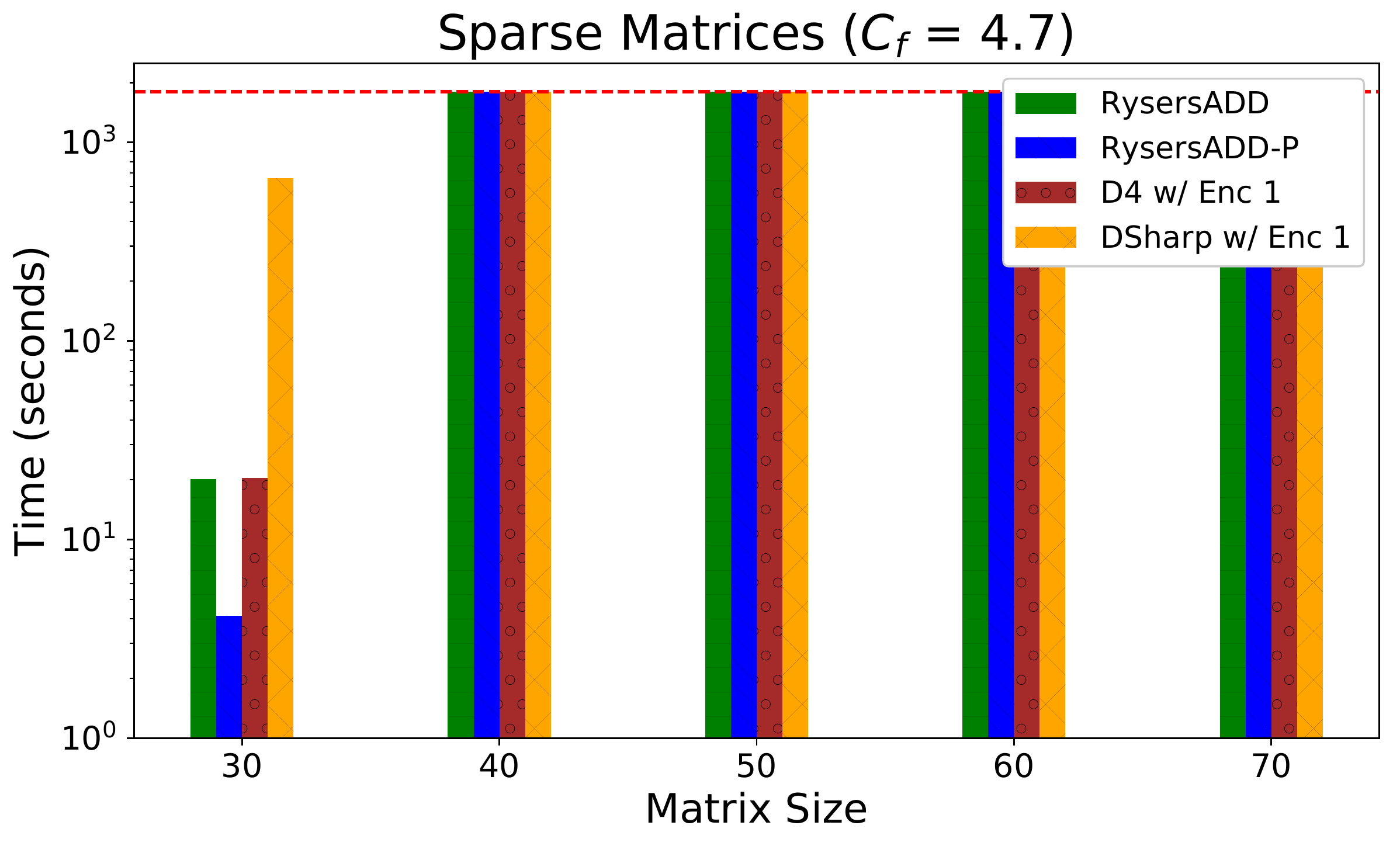}
	\end{subfigure}
	\hfill
	\begin{subfigure}{0.48\textwidth}
		
		\includegraphics[width=\textwidth]{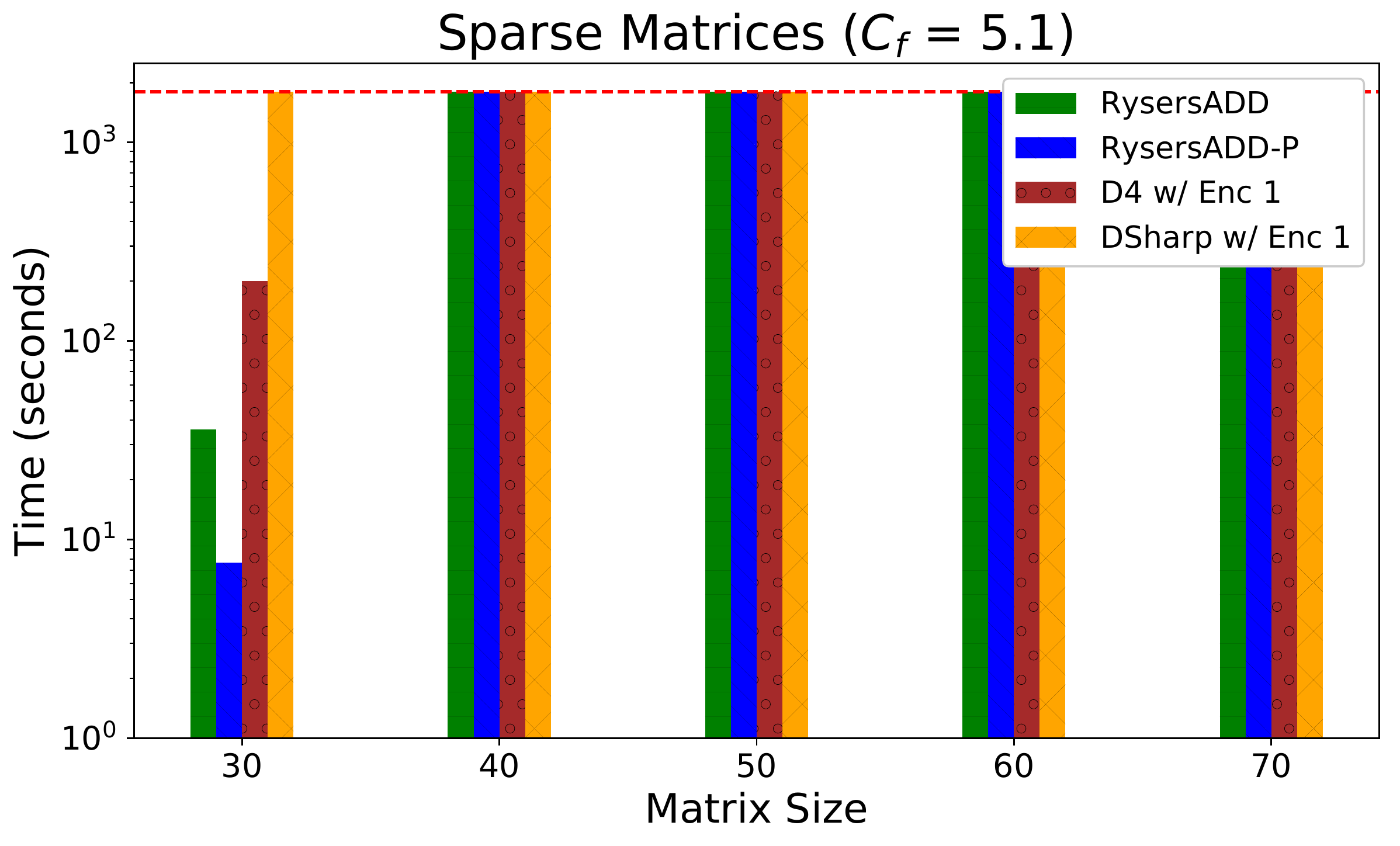}
	\end{subfigure}
	\caption{Performance on Sparse Matrices}
	\label{fig:sparse}
\end{figure}
\subsection{Performance on sparse matrices}
Fig. \ref{fig:sparse} depicts the median running times of the algorithms for sparse matrices with $ C_f \in \{3.9,4.3,4.7,5.1\} $. We plot the running time of the ADD-based approaches with early abstraction (see Sec. \ref{sec:algs}). Monolithic variants (not shown) time out on all instances with $ n \ge 40 $. For {\dfr} and {\ds}, we plot the running times only for Pairwise encoding of At-Most-One constraints, since our
preliminary experiments showed that it substantially outperformed
other encodings. We see that {\dfr} is the fastest when sparsity is high i.e. for $ C_f \le 4.3 $, but for $ C_f \ge 4.7 $ the ADD-based methods are the best performers. {\ds} is outperformed by the remaining $3$ algorithms in general.

\begin{figure}
	\centering
	\begin{subfigure}{0.48\textwidth}
		\includegraphics[width=\textwidth]{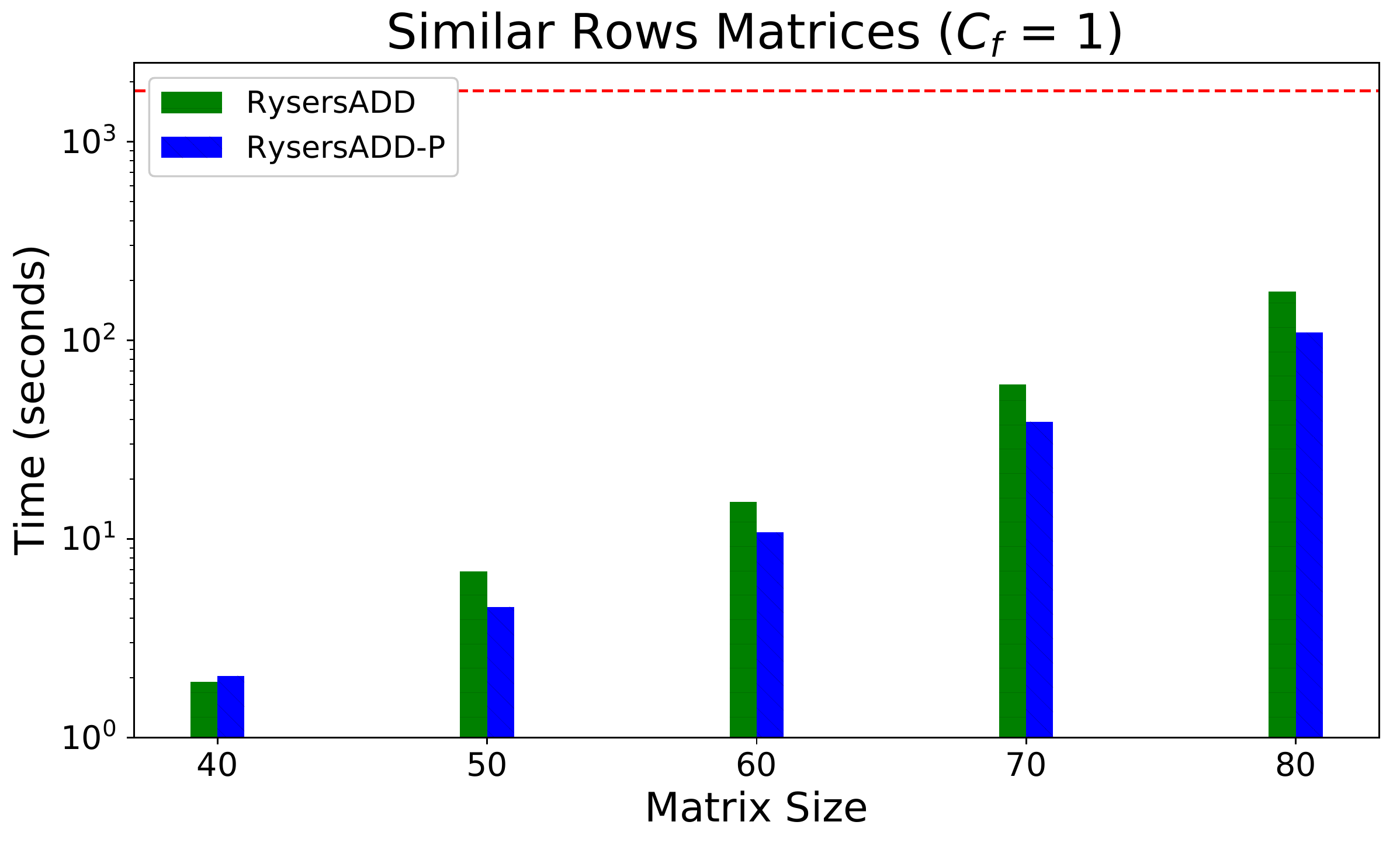}
	\end{subfigure}
	\hfill
	\begin{subfigure}{0.48\textwidth}
		
		\includegraphics[width=\textwidth]{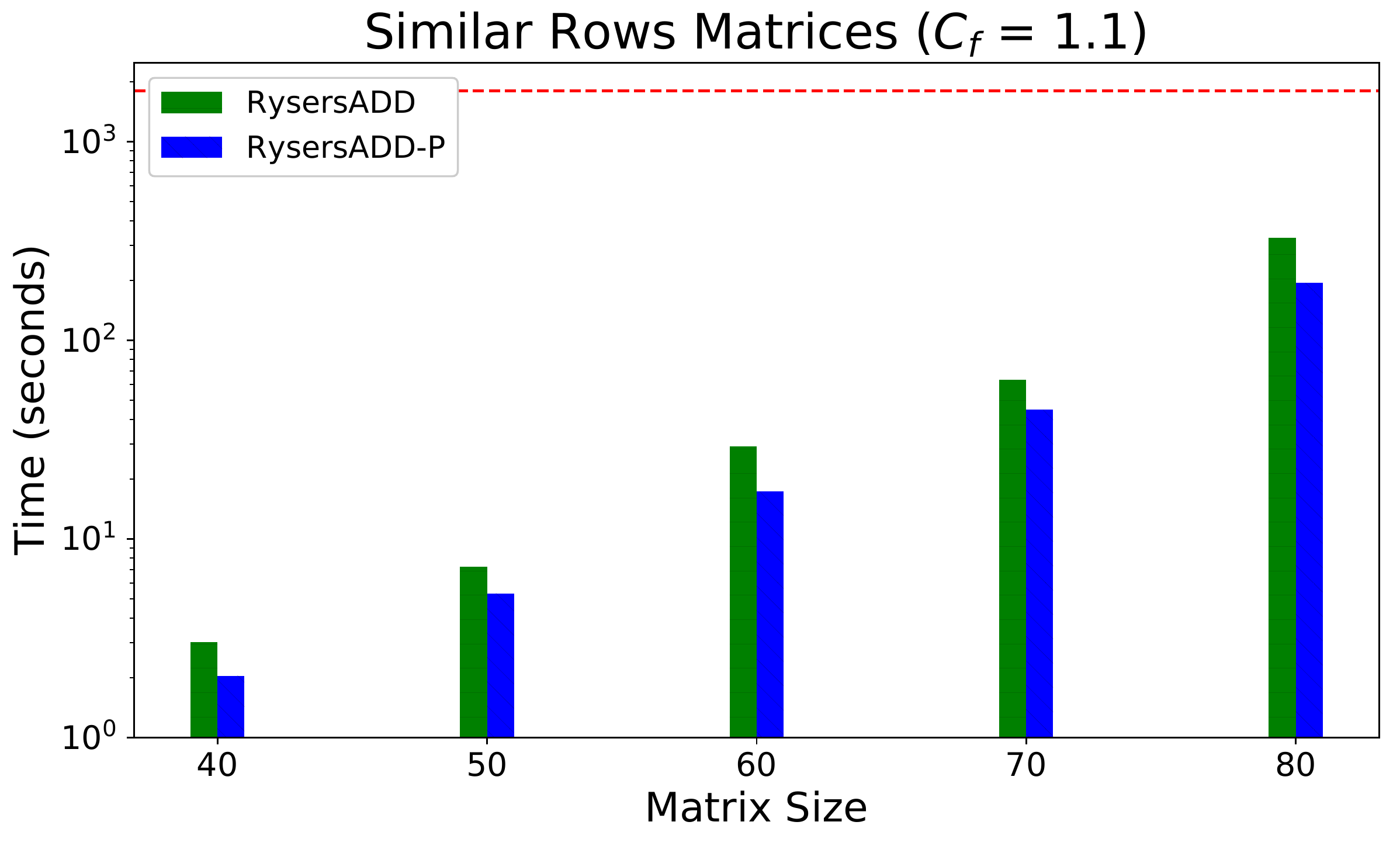}
	\end{subfigure}
	\begin{subfigure}{0.48\textwidth}
		\includegraphics[width=\textwidth]{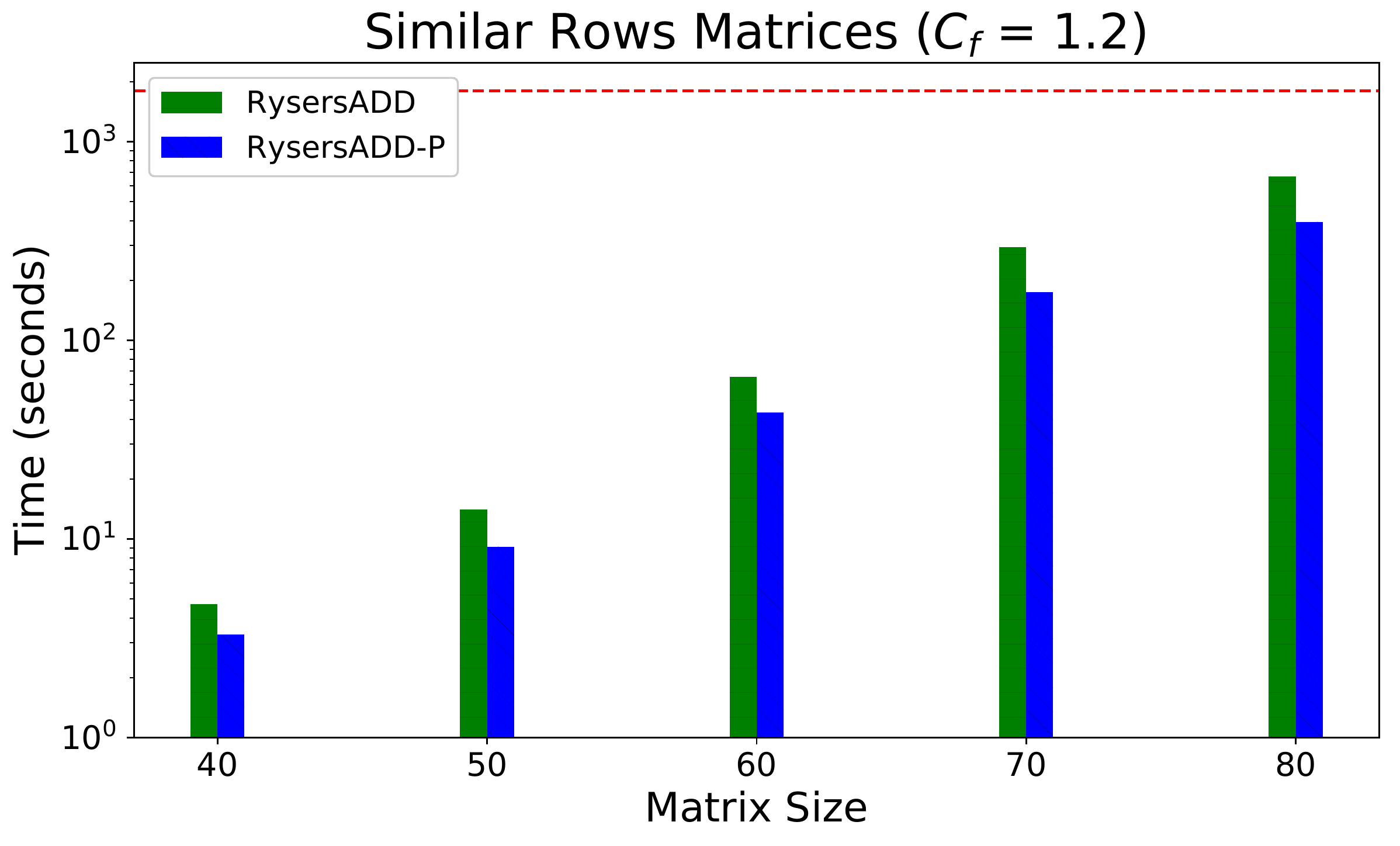}
	\end{subfigure}
	\hfill
	\begin{subfigure}{0.48\textwidth}
		
		\includegraphics[width=\textwidth]{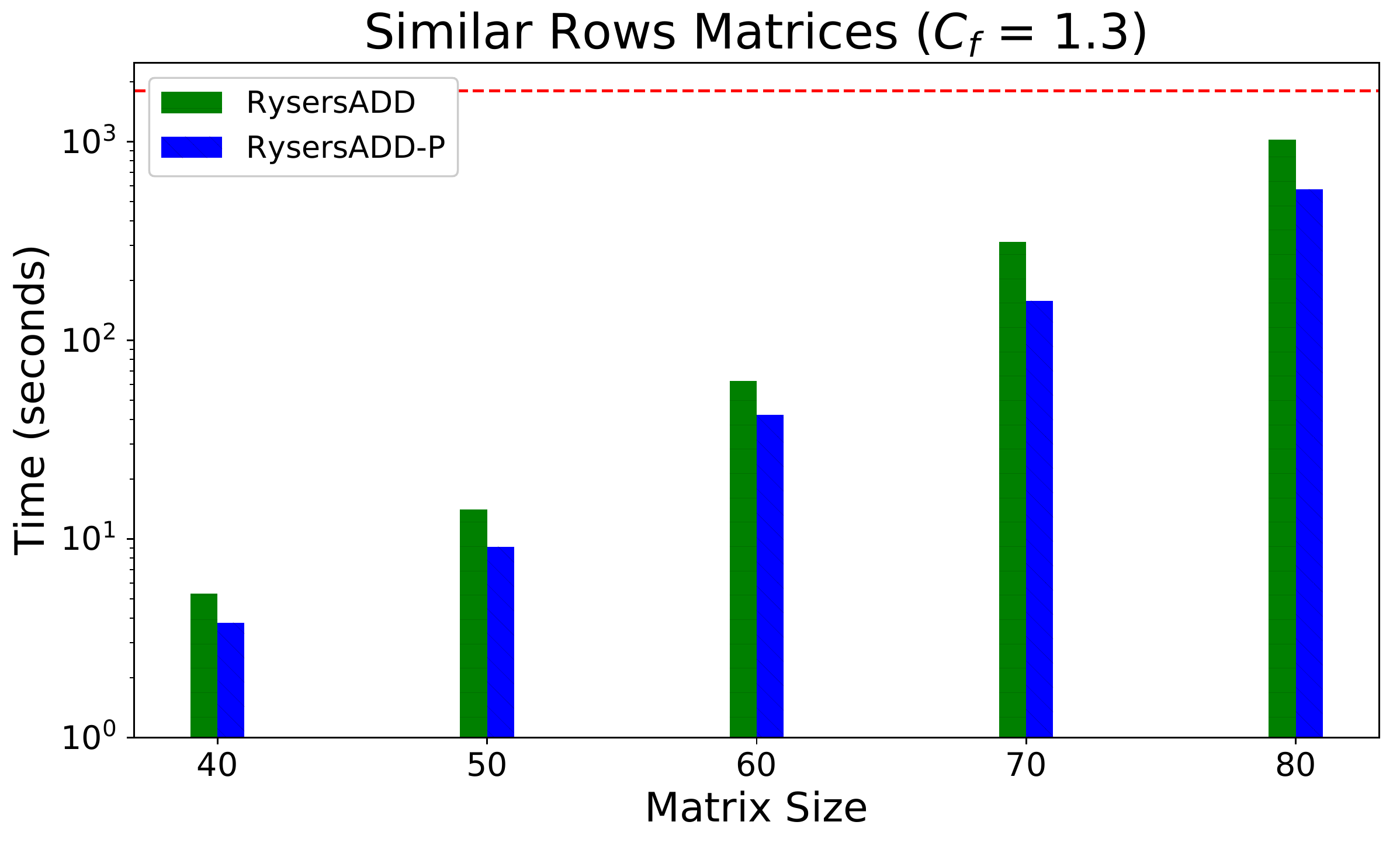}
	\end{subfigure}
	\caption{Performance on similar-rows matrices. {\dfr}, {\ds} (not shown) timeout on all instances.}
	\label{fig:similar}
	\vspace{-0.5cm}
\end{figure}
\subsection{Performance on similar-row matrices}
Fig. \ref{fig:similar} shows plots of the median running time on similar-row
matrices with $ C_f = \{1,1.1,1.2,1.3\} $. We only present the case
when $ \rho = 0.8 $, since the plots are similar when $ \rho \in
\{0.7, 0.9\} $. As in the case of dense matrices, {\dfr} and
   {\ds} were unable to solve any instance of size $40$, and hence we
   only show plots for {\radd} and {\raddp}. The performance of both
   tools is markedly better than in the case of dense matrices, and
   they scale to matrices of size $80$ within the $1800$
   second timeout.

\begin{figure}[b]
	\centering
	\includegraphics[width=3in]{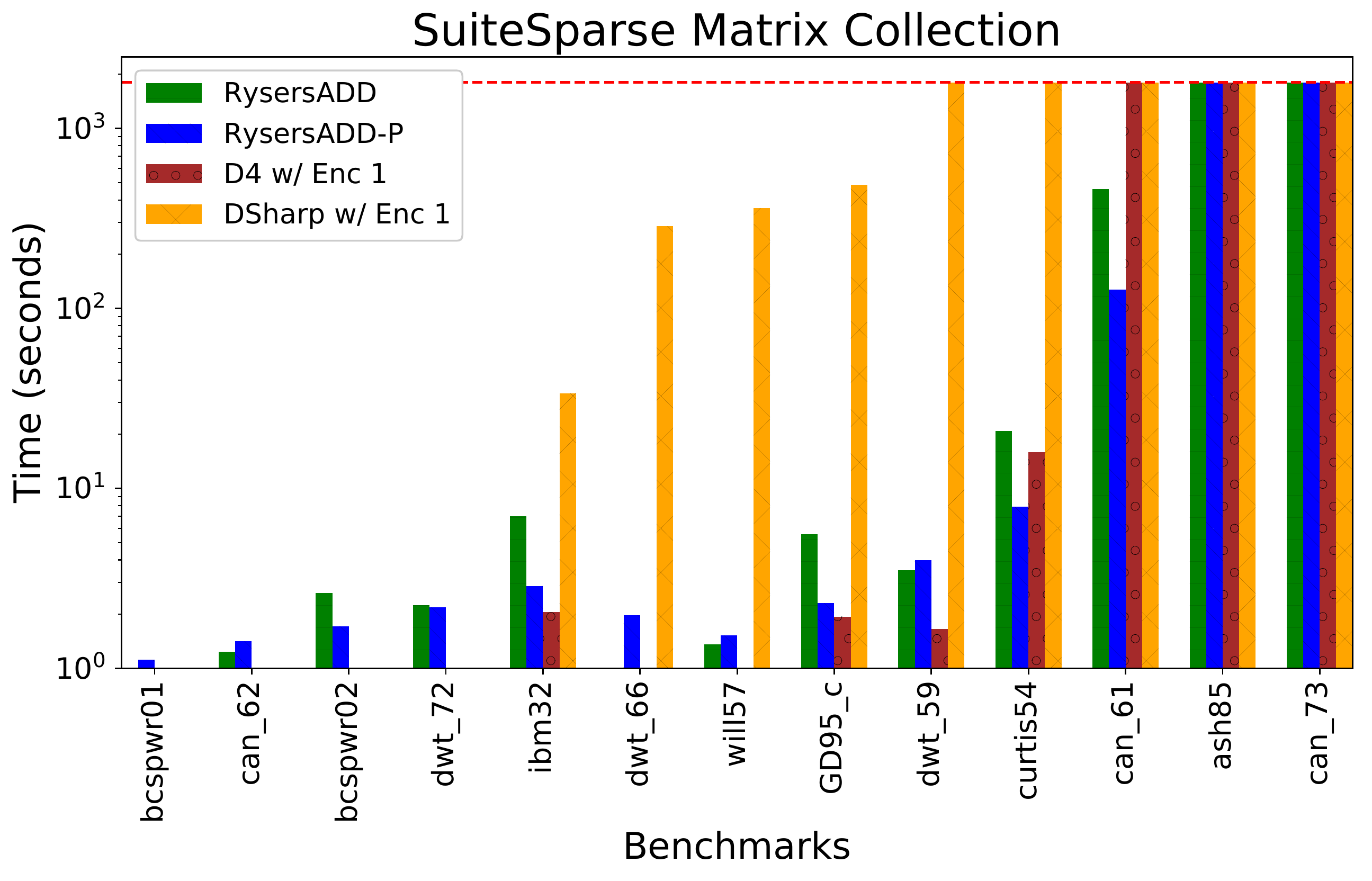}
	\caption{}
	\label{fig:tamu_sparse}
	\vspace{-0.5cm}
\end{figure}
\subsection{Performance on SuiteSparse Matrix Collection}
We report the performance of algorithms $\radd$, $\raddp$, {\dfr} and
  {\ds} on $13$ representative graphs from the SuiteSparse Matrix
  Collection in Fig. \ref{fig:tamu_sparse}. Except for the first $4$
  instances, which can be solved in under 5 seconds by all algorithms,
  we find that {\dfr} is the fastest in general, while the
  ADD-based algorithms outperform {\ds}. Notably, on the instance
  "can\_61", both {\dfr} and {\ds} time out while {\radd} and {\raddp}
  solve it comfortably within the alloted time. We note that the
  instance "can\_61" has roughly $9n$ $1$s, while {\dfr} is the best
  performer on instances where the count of $1$s in the matrix lies
  between  $4n$ and $6n$.

\begin{table}
	\vspace{-0.8cm}
	\caption{Running Times on the fullerene $ C_{60} $. EA: Early Abstraction Mono: Monolithic}
	\resizebox{\textwidth}{!}{%
		\begin{tabular}{|c|c|c|c|c|c|c|c|c|c|c|c|c|c|c|c|c|}
			\hline
			\textbf{Tool}                                             & \multicolumn{6}{c|}{\textbf{\dfr}}     & \multicolumn{6}{c|}{\textbf{\ds}} & \multicolumn{2}{c|}{\textbf{{\radd}}} & \multicolumn{2}{c|}{\textbf{{\raddp}}} \\ \hline
			\begin{tabular}[c]{@{}c@{}}\textbf{Encoding /}\\ \textbf{Mode}\end{tabular} & 1    & 2     & 3     & 4   & 5   & 6   & 1   & 2   & 3   & 4   & 5   & 6   & EA                & Mono              & EA                 & Mono              \\ \hline
			Time (sec)                                                & 94.8 & 150.5 & 150.6 & 136 & 158 & 156 & \multicolumn{6}{c|}{TimeOut}      & 96.4              & TimeOut              & 57.1               & TimeOut              \\ \hline
		\end{tabular}%
	}
	\label{tab:ful}
\end{table}
\subsection{Performance on fullerene adjacency matrices}
We compared the performance of the algorithms on the adjacency matrices of the fullerenes $ C_{60} $ and $
C_{100} $.  All the algorithms timed out on $ C_{100} $. The results
for $ C_{60} $ are shown in Table~\ref{tab:ful}.  The columns under {\dfr} and {\ds} correspond to
$6$ different encodings of At-Most-One constraints (see Sec. \ref{sec:algs}). It can be
seen that {\raddp} performs the best on this class of matrices,
followed by {\dfr}. The utility of early abstraction is clearly
evident, as the monolithic approach times out in both cases.

\vspace{0.3cm}
\noindent\textbf{Discussion}: Our experiments show the
effectiveness of the symbolic approach on dense
and similar-rows matrices, where neither {\dfr} nor {\ds} are able to
solve even a single instance. Even for sparse
matrices, we see that decreasing sparsity has lesser effect on the
performance of ADD-based approaches as compared to {\dfr}. This trend
is confirmed by "can\_61" in the SuiteSparse Matrix Collection as
well, where despite the density of $1$s being $ 9n $, {\radd} and
{\raddp} finish well within timeout, unlike {\dfr}.  In the case of
fullerenes, we note that the algorithm in \cite{liang2004partially}
solved $ C_{60} $ in 355 seconds while the one in
\cite{yue2013improved} took 5 seconds, which are in the vicinity of
the times reported in Table~\ref{tab:ful}. While this is not an
apples-to-apples comparison owing to differences in the computing
platform, it indicates that the performance of general-purpose
algorithms like {\radd} and {\dfr} can be comparable to that of 
application-specific algorithms.

%% file: Conclusion.tex
\section{Conclusion}

In this work we introduced a symbolic algorithm called {\radd} for
permanent computation based on augmenting Ryser's formula with
Algebraic Decision Diagrams. We demonstrated, through rigorous
experimental evaluation, the scalability of {\radd} on both dense and
similar-rows matrices, where existing approaches fail. Coupled with
the technique of early abstraction~\cite{addmc}, {\radd} performs
reasonably well even on sparse matrices as compared to dedicated
approaches. In fact, it may be possible to optimize the algorithm
even further, by evaluating other heuristics used in~\cite{addmc}.  We leave this for future work.  Our work also re-emphasizes the
versatility of ADDs and opens the door for their application to other combinatorial problems.

It is an interesting open problem to obtain a complete
characterization of the class of matrices for which ADD representation
of Ryser's formula is succinct. Our experimental results for dense
matrices hint at the possibility of improved theoretical bounds
similar to those obtained in earlier work on sparse matrices.
Developing an algorithm for general matrices that is exponentially
faster than Ryser's approach remains a
long-standing open problem~\cite{izumi2012new}, and obtaining better bounds for non-sparse matrices would be an important first step in this direction.

